\documentclass[aps,pra,a4paper,10pt,twocolumn,nofootinbib]{revtex4-2} 
\usepackage[T1]{fontenc}
\usepackage{mathptmx}
\usepackage{datetime}
\usepackage{amsmath}
\usepackage{amsfonts}
\usepackage{amsfonts}
\usepackage{mathrsfs}
\usepackage[mathscr]{euscript}
\usepackage[dvips]{graphicx}
\usepackage{fancyhdr}
\usepackage{colordvi}
\usepackage{hyperref} 
\usepackage{epsfig}
\usepackage{color}
\usepackage{bm}
\usepackage{enumitem}

\usepackage{float}
\def\overstrike#1#2{{\setbox0\hbox{$#2$}\hbox to \wd0{\hss
    $#1$\hss}\kern-\wd0\box0}}

\newdateformat{yymmdddate}{\THEYEAR/\twodigit{\THEMONTH}/\twodigit{\THEDAY}}

\def\DS{Dst}

\def\KP{K_\textup{P}}

\def\DKL{D_\textup{KL}}
\def\SKL{S_\textup{KL}}

\def\doff{d_D}
\def\scal{s_D}

\begin{document}
\title{The ${\DS}$ index -- empirical fits and temporal correlations}
\author{Paul Kinsler}
\homepage[]{https://orcid.org/0000-0001-5744-8146}
\affiliation{
  Department of Electronic \& Electrical Engineering, 
  University of Bath,
  Bath,
  BA2 7AY,
  United Kingdom
}

\lhead{\includegraphics[height=5mm,angle=0]{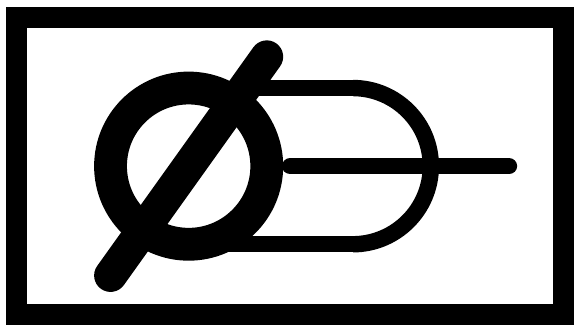}~~FITADAST}
\chead{${\DS}$ -- fits and correlations}
\rhead{
\href{mailto:Dr.Paul.Kinsler@physics.org}{Dr.Paul.Kinsler@physics.org}
}

\begin{abstract}

Here I show how it is possible
 to make an empirical fit of 
 yearly sets of ${\DS}$ index values
 to the log-normal distribution, 
 using the Kullback-Liebler divergence 
 to test each fit performance.
Further, 
 since any real sequence of ${\DS}$ values
 contains correlation with the near-past, 
 I also compute autocorrelations
 in order to understand the timescales over which
 it retains memory.

\end{abstract}


\date{\today}
\maketitle
\thispagestyle{fancy}

%

%
\section{Introduction}\label{S-introduction}

The disturbance storm time index (${\DS}$) 
 \cite{Nose-ISK-2015}
 indicates the size of the Earth's ring current, 
 i.e. the current that circulates around the Earth.
The ring current produces a magnetic field that is opposes the Earth's magnetic field, 
 and it is this field alteration which $\DS$ actually corresponds to. 
However, 
 ${\DS}$ dynamics also depends on other magnetospheric current systems,
 such as magnetopause and tail currents
 \cite{Liemohn-KTELBC-2001sp,Vasyliunas-2006sp}.
These dynamics play a role in the Earth's magnetosphere \cite{Borovsky-V-2018sg},
 and a negative ${\DS}$ value means that the magnetic field 
 around the Earth is weakened, as can be the case during 
 significant space weather events \cite{Ishii-BFHBR-2024asr}.
Notably, 
 the ${\DS}$ index has long been used as a key parameter
 informing space weather research \cite{Sugiura-1964aigy,Nose-ISK-2015}
 as well as being an indicator of geomagnetic storm intensity.

Here
 I show how 
 it is possible to fit the distribution(s) of ${\DS}$ values
 within some time period to log-normal distributions.
However, 
 key to this process is 
 is not only using a year long ${\DS}$ sequence so as to
 average the rather variable and distinctly intermittent solar drivers,
 but also adding an offset ${\doff}$ to the ${\DS}$  values 
 before fitting the ${\DS}$  histogram.

Here I fit year-long slices taken from the 
 67-years worth of ${\DS}$ values 
 that can be obtained from 
 the WDC for Geomagnetism service (https://wdc.kugi.kyoto-u.ac.jp).
As used here, 
 the downloaded data spans from 1957 up until June 2025.
One consequence of the fits and of autocorrelation calulations 
 is that it enables the creation of a simple method for 
 creating sequences
 of synthetic-${\DS}$ values; 
 sequences that respect both the distribution
 and the memory timescales of actual ${\DS}$ time series.
Although no substitute for physically motivated predictions of ${\DS}$
 (see e.g. \cite{Burton-MR-1975sp,Wang-CL-2003sp}), 
 such sequences might have value in providing
 of reasonably plausible synthetic drivers to test 
 downstream models of (e.g.) the ionosphere.

Systematic attempts to fit ${\DS}$ distributions
 over its whole history from 1957 to the present 
 do not appear to be present in existing literature; 
 and so 
 the approach here deliberately parallels the
 $\KP$ fits calculated elsewhere \cite{Kinsler-2025draft-fitakarp}.

In what follows I discuss the choice of the log-normal distribution
 in Section \ref{S-why}
 before addressing the fits themselves in Section \ref{S-data}, 
 and temporal correlations 
 and the simple synthetic time-series construction in
 Section \ref{S-autocorrelations}.
Finally,
 in Section \ref{S-summary} I conclude, 
 leaving an exhaustive set of figures
 showing the yearly fits for Appendix \ref{S-yearfits}
 and the fit parameters themselves for Appendix \ref{S-fitparams}.

\section{Why log-normal?}\label{S-why}

On the face of it, 
 the log-normal distribution is an unsuitable candidate
 for fitting ${\DS}$ distributions, 
 since ${\DS}$ is in-principle unbounded, 
 and log-normal (LN) distributions are bounded at zero.
Here, 
 adding a paameterised offset 
 avoids the fact that log-normals are bounded at zero
 by shifting the data away from the new origin.

However, 
 the choice here is a pragmatic one
 and the LN turns out to have
 a peaked form with
 a compatible degree of asymmetry.
Other distributions were tried, 
 again motivated by achieving a pragmatic fit 
 rather than due to a motivation 
 derived from physical principles.
The logistic function performed similarly to the LN, 
 but the Gamma was generally too wide
 (and so not peaked enough),
 and the Weibull was even worse in this regard.
And of course, 
 the usual normal distribution lacked any asymmetry at all.

Although ${\DS}$ values are not bounded 
 in any simple way, 
 they nevertheless usually
 lie within within a typical range,
 whilst
 varying due to the heliospheric drivers of the magnetosphere
 as well as the local physical processes.
The tails of the ${\DS}$ distribution are (or can be) wide, 
 but those tails have minimal probability,
 meaning that the boundedness of the LN distribution
 causes no particular issues 
 with respect to achieving a good fit to 
 the typical range of the ${\DS}$ distributions.

\section{Fitting ${\DS}$}\label{S-data}

\begin{figure}
\includegraphics[width=0.490\columnwidth]{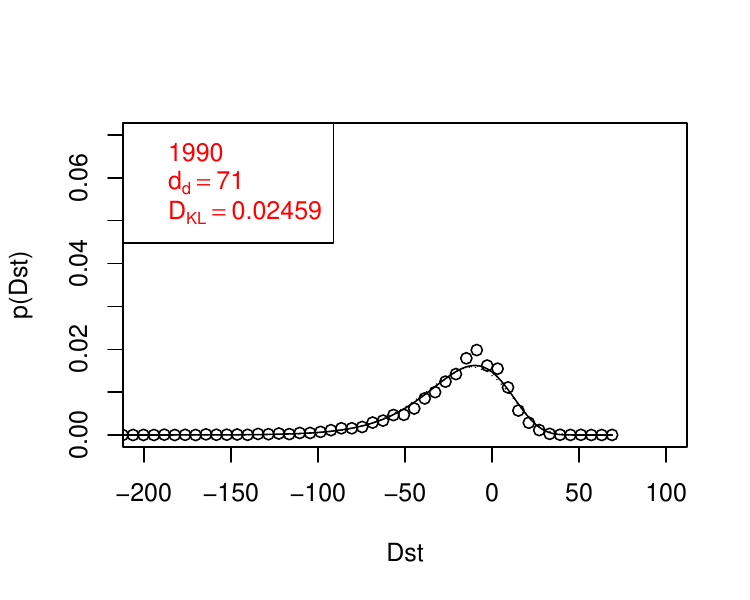}
\includegraphics[width=0.490\columnwidth]{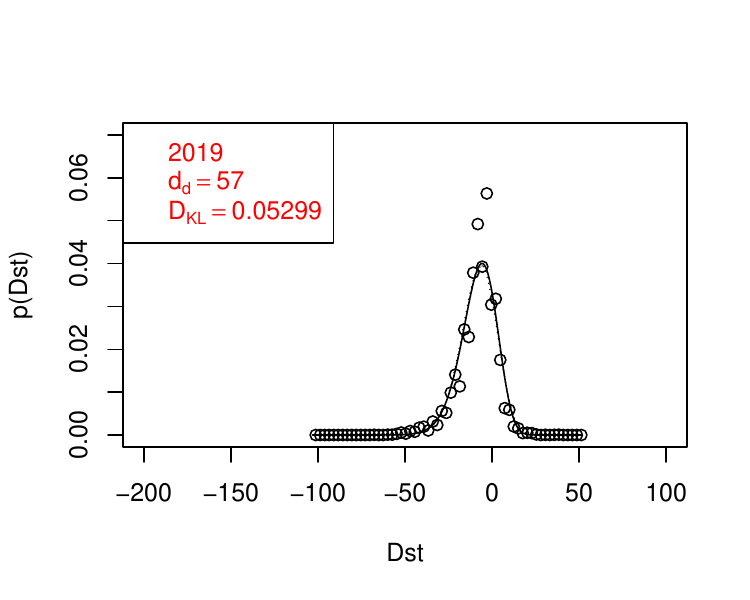}
\caption{Fitting results for two contrasting sample years, 
 1990 (active, with a broader distribution) 
 and 2019 (quiet, with a narrow distribution).
The circles indicate the ${\DS}$ histogram, 
 the solid line indicates the fitted distribution.
Here both legends are coloured red to 
 highlight that the KL divergence 
 indicates a likely statistical mismatch between 
 the fit and the data.
A full set of figures is given in Appendix \ref{S-yearfits}.
}
\label{fig-samples}
\end{figure}

\def\scal{s_D}

Here I do not apply a simple fit of the LN distribution
 directly to ${\DS}$ histograms, 
 a process that could only give unsatisfactory results.
Therefore I
 vary the offset of the ${\DS}$ data
 before attempting a fit.
That is,
 for some collection of ${\DS}$  values
 I generate a ${\DS}'$ such that
~
\begin{align}
  {\DS}' 
&=
    {\doff} - {\DS}
\end{align}
~
and iterate over ${\doff}$ 
 to find the best log-normal match to the histogram
 of ${\DS}'$ values.
Here, 
 I scanned
 ${\doff}$ values from $40$nT to $140$nT; 
 then dividing the resulting range into 60 bins
 to create a fittable histogram of $\DS$ values.
I then chose the best fit as that 
 which had the smallest Kullback-Liebler (KL) divergence ${\DKL}$ \cite{Kullback-Leibler,Kullback-L-1951ams}
 between the data and the fit.
Here I do the 
 fits using the \textsf{MASS} library \cite{R-MASS}, 
 along with the \textsf{philentropy} library for computing  KL divergences \cite{R-philentropy}; 
 both for the R language \cite{R-language}. 
I use the finalised $\DS$ data for years 1957--2020,
 and the provisional data for 2021 onwards; 
 according to what was available at the time of writing.

Once a best fit has been obtained for a data set, 
 we can then compute its properties --
 here the mode (peak position) and standard deviation
 in terms of the ${\DS}'$ scale; 
 and then revert back to the original ${\DS}$ scale.

\begin{figure}
\includegraphics[width=0.990\columnwidth]{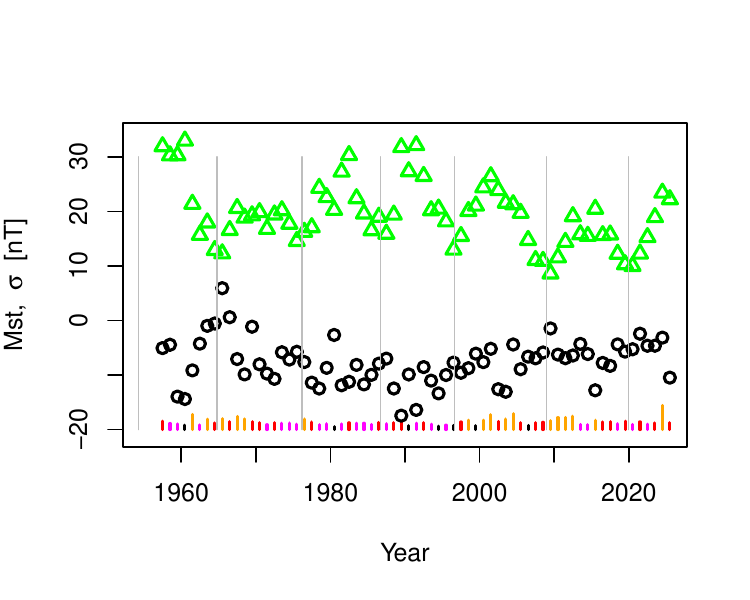}
\caption{Fitting results by year: 
 The  black circles indicate the peak position
  ($Mst$, the mode)
  of the fitted log-normal distribution, 
 and the green triangles indicate the standard deviation ($\sigma$).
The bars at the bottom indicate ${\DKL}$, 
 and in particular whether it is less than $2{\SKL}$ (black),
 between $2{\SKL}$ and $3{\SKL}$ (magenta), 
 between $3{\SKL}$ and $4{\SKL}$ (red), 
 or greater than $4{\SKL}$ (orange).
Vertical grey lines indicate 
 the start of each solar cycle.
}
\label{fig-summary}
\end{figure}

Sample outcomes from this process 
 are given on fig. \ref{fig-samples}, 
 and the overall outcomes 
 are summarised on fig. \ref{fig-summary}.
These results are somewhat ``noisy'' -- 
 i.e. they do not suggest a smooth progression
 from year to year, 
 but presumably such might be achieved by 
 doing fits on year-long sliding window
 rather than just on disjoint year data.

\begin{figure}
\includegraphics[width=0.990\columnwidth]{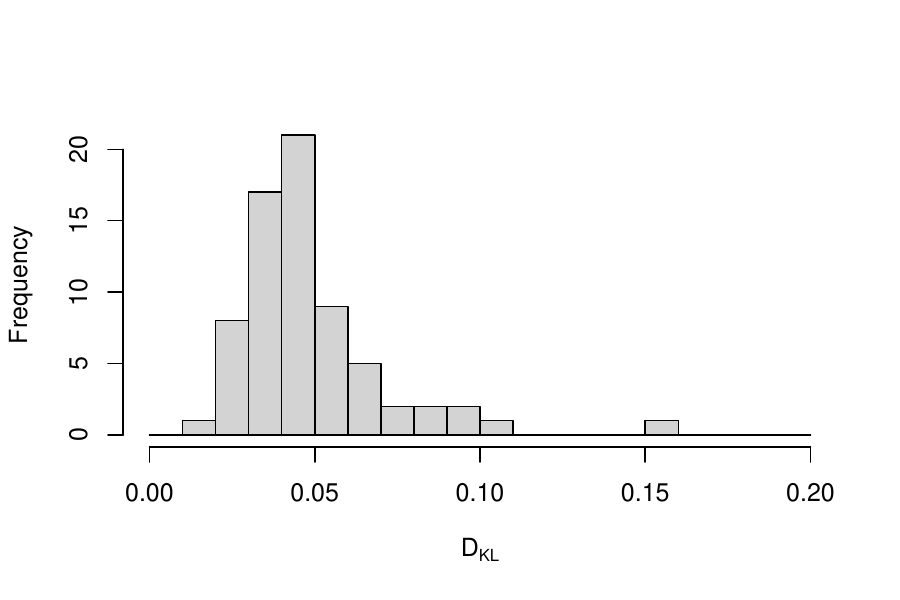}
\caption{Histogram of the KL divergences accumulated 
 over each year of the dataset -- 
 we can see that most are less than $0.10$. 
Given the sampling density, 
 this indicates that whilst the fits to a LN distribution
 perform quite well, 
 the match is not exact.
}
\label{fig-KLDhisto}
\end{figure}

Generally, 
 and as can be seen in section \ref{S-yearfits}, 
 the fitted distributions are usually an encouraging --
 albeit not perfect -- 
 visual match to the data.
Indeed, 
 as reported on the figure legends, 
 and summarized for the full dataset in fig. \ref{fig-KLDhisto},
 and the tables in section \ref{S-fitparams},
 the KL divergences \cite{Kullback-Leibler,Kullback-L-1951ams}
 are reasonably small.

A small divergence indicates not only a good fit, 
 but a reliable indication of the goodness of fit, 
 due to the dense sampling of the full ${\DS}$ range --
 in a non leap year, 
 we have $24 \times 365 = 8760$ values
 filling 60 histogram bins, 
 for an average bin-population of $146$.
The expected KL divergence for sampling error at our sampling density of $146$/bin
 is  ${\SKL} = 1/146 \simeq 0.007$ \cite{Kinsler-2026zenodo-kullbex};
 so any value significantly larger than that 
 should therefore be due to shortcomings in the LN model
 for that year's data.
Notably, 
 all the histogram fits fail the ${\DKL} \le {\SKL}$ test.
This means that although by eye they might adequately represent
 the basic location and width properties of the ${\DS}$ distributions, 
 the amount of data available should --
 if a LN were the appropriate distribution --
 give a closer fit.

\begin{figure}
\includegraphics[width=0.990\columnwidth]{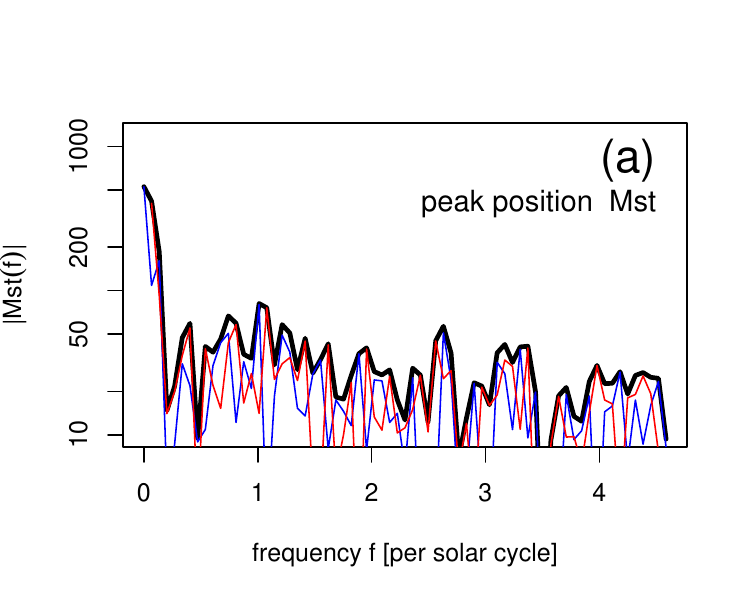}
\includegraphics[width=0.990\columnwidth]{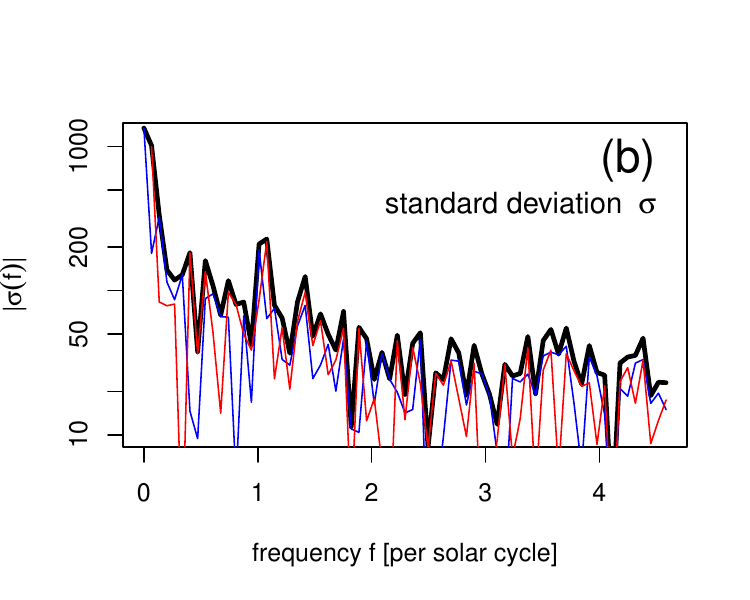}
\caption{Fourier transform 
 of the year-by-year fitted-peak position (a)
 and the standard deviation (b).
The black line shows the 
 absolute value of the Fourier transform, 
 the blue and red are the same for the real and the imaginary parts.
}
\label{fig-fourier}
\end{figure}

Fourier transforms of the fitting results were also calculated
 after first zero-padding the data from 1890 
 until the data start at 1957.
Unsurprisingly, 
 these
 indicate a solar cycle period 
 in keeping with the accepted values, 
 as seen on fig. \ref{fig-fourier}; 
 this periodicity is also apparent in the standard deviation
 time series on fig. \ref{fig-summary}.

Of course it is possible to 
 derive a standard deviation and a peak position 
 directly from the $\DS$ data, 
 with no fitting necessary.
Nevertheless, 
 that would not guarantee that the those parameters
 thus extracted were sufficient to parameterise the 
 distribution; 
 it might still be that other parameters would be needed, 
 e.g. such as higher-order moments.
However, 
 the success of the fitting process
 indicates that a reasonable three-parameter
 fit to annual ${\DS}$ distributions is possible
 (namely the offset ${\doff}$,
  and the two LN parameters 
  $\mu$ and $\sigma$).

Notwithstanding the broad success of this fitting process, 
 it is also true that the larger than ideal divergences ${\DKL}$ and 
 occasional anomalous distribution suggests that 
 there may be more to be discovered.
This might be partly due to the fact that hour-by-hour $\DS$ variation
 is highly contingent on circumstances, 
 in particular as regards excursions due to 
 the intermittent arrival of CMEs.
This intermittency means that the fits can only be applied 
 to long sequences of data -- here years -- 
 since (e.g.) monthly distributions are much less well behaved.
Also, 
 in active years, 
 with their wider distributions, 
 and if instead plotted on a logarithmic scale,
 the LN sometimes falls off too fast for the data at large negative $\DS$; 
 and although this does not affect the linear-scale visual fit, 
 does affect the ${\DKL}$.
This deficiency suggests that the LN does not always properly 
 represent the statistics of relatively rare ``storm time'' events,
 where $\DS$ is driven more strongly negative;
 although it may be rather because consistent sampling of such events
 cannot always be achieved on only year timescales.

\section{Temporal correlations}\label{S-autocorrelations}

One might imagine trying to use these distributions as a mechanism
 to generate a plausible synthetic time-series of $\DS$ values, 
 perhaps as a test driver of some ionospheric model or other system 
 influenced by global properties of the magnetosphere.
However, 
 these bare distributions --
 or similar ones abstracted from a non-calendar year of $\DS$ data --
 are not by themselves immediately suitable; 
 they do not tell us how subsequent values of $\DS$ will be 
 related to one another; 
 sampling directly from a selected distribution
 does not respect the correlations
 that exist between subsequent ${\DS}$ values in real data.
Here 
 I partly address this by allowing for
 an exponential decay in correlations,
 although more sophisticated procedures
 might be imagined.

To create a LN distributed time series we can simply exponentiate 
 a normally-distributed time series; 
 but here we need to ensure that the 
 coefficients of the normally-distributed process 
 are chosen to match the LN distributed properties of $\DS$.
Therefore
 in a scenario with pre-fitted (or pre-chosen)
 parameters, 
 we can generate a normally-distributed time series
 with memory timescale $T$
 using the loss rate  $\gamma = \exp(-1\textrm{hr}/T)$ and 
 standard deviation $\sigma$.
Conveniently, 
 since the fits are done using the MASS library \cite{R-MASS}
 for the R language \cite{R-language}, 
 the routines provide $\mu$ and $\sigma$ directly.
This means that 
 along with the fit-optimisation that works out $\doff$ and $\scal$
 (see table),
 conversion from LN properties is not required.
We can therefore use 
 the update equation 
~
\begin{align}
  Z(t+1hr)
&=
  \mu
 +
  \left(1 - \gamma \right)
  \left( Z(t) - \mu \right)
 +
  {\sigma}\sqrt{2\gamma} W(t)
,
\label{eqn-Zdiffusion}
\end{align}
 where $W(t)$ is a gaussian noise term with 
 zero average and unit standard deviation.
This directly gives us a 
 LN time series with the required properties as
~
\begin{align}
  {\DS}'(t)
&=
  \exp \left[ Z(t) \right]
\label{eqn-dstcorrelates}
\\
\textrm{or}\qquad
  {\DS}(t)
&=
  \left\{ \doff - \scal \exp \left[ Z(t) \right] \right\}
,
\label{eqn-dstconversion}
\end{align}
 which have 
 mean $M_{\DS} = \exp(\mu + \sigma^2/2)$
 and variance $V_{\DS} = \exp(\sigma^2-1) \exp(2\mu+\sigma^2)$.

\begin{figure}
\includegraphics[width=0.990\columnwidth]{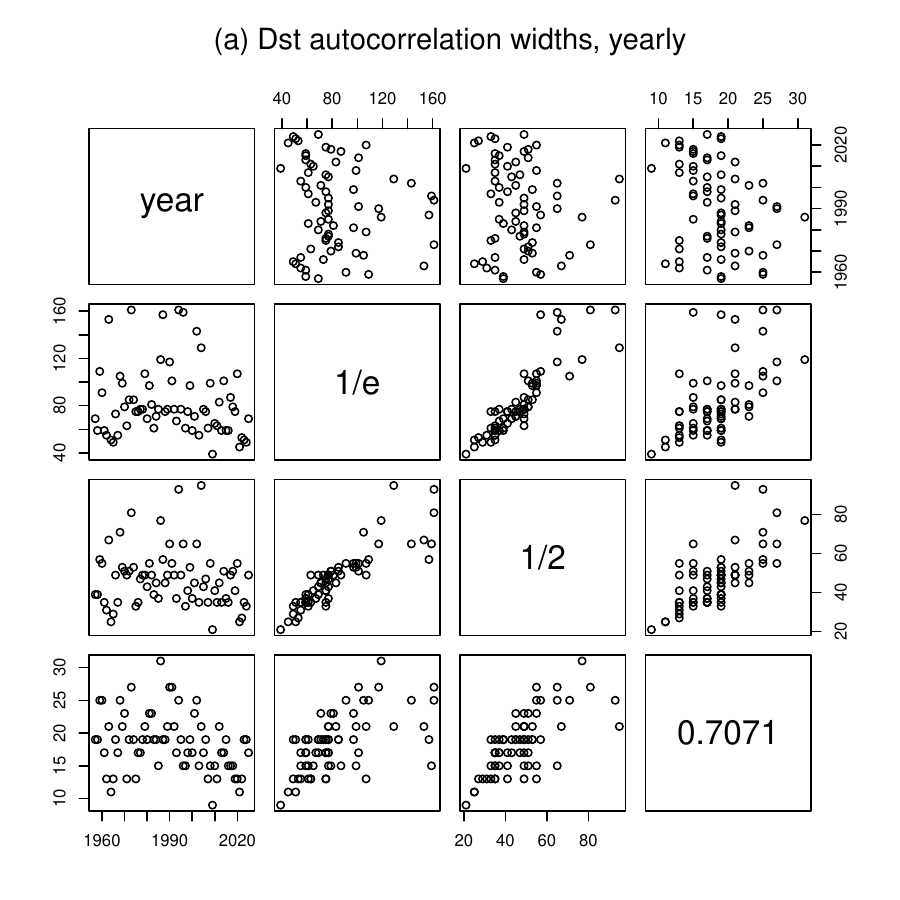}
\includegraphics[width=0.990\columnwidth]{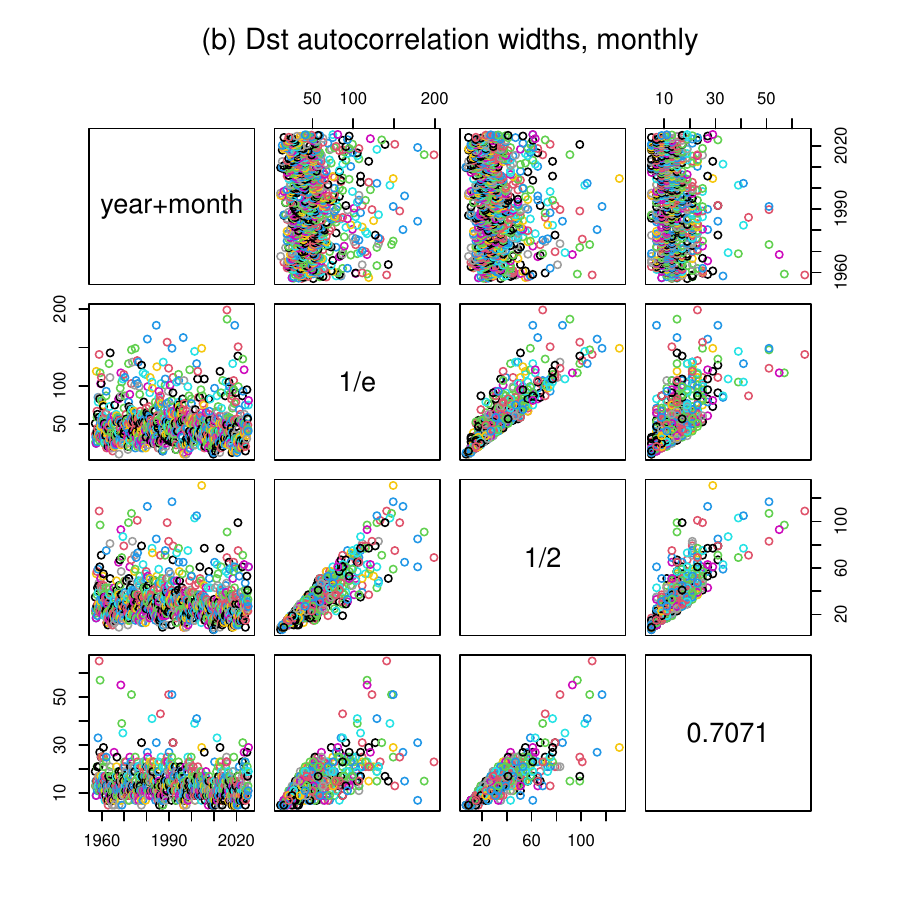}
\caption{Scatter plots of 
 Autocorrelation function widths 
 on time-series subsets (yearly and monthly) 
 for ${\DS}$ values.
Widths (full widths) are calculated at threshold values of 
 $1/e$, $1/2$, and $1/\sqrt{2} \simeq 0.7071$
 of the maximum.
}
\label{fig-autocorr}
\end{figure}

\begin{figure}
\includegraphics[width=0.790\columnwidth]{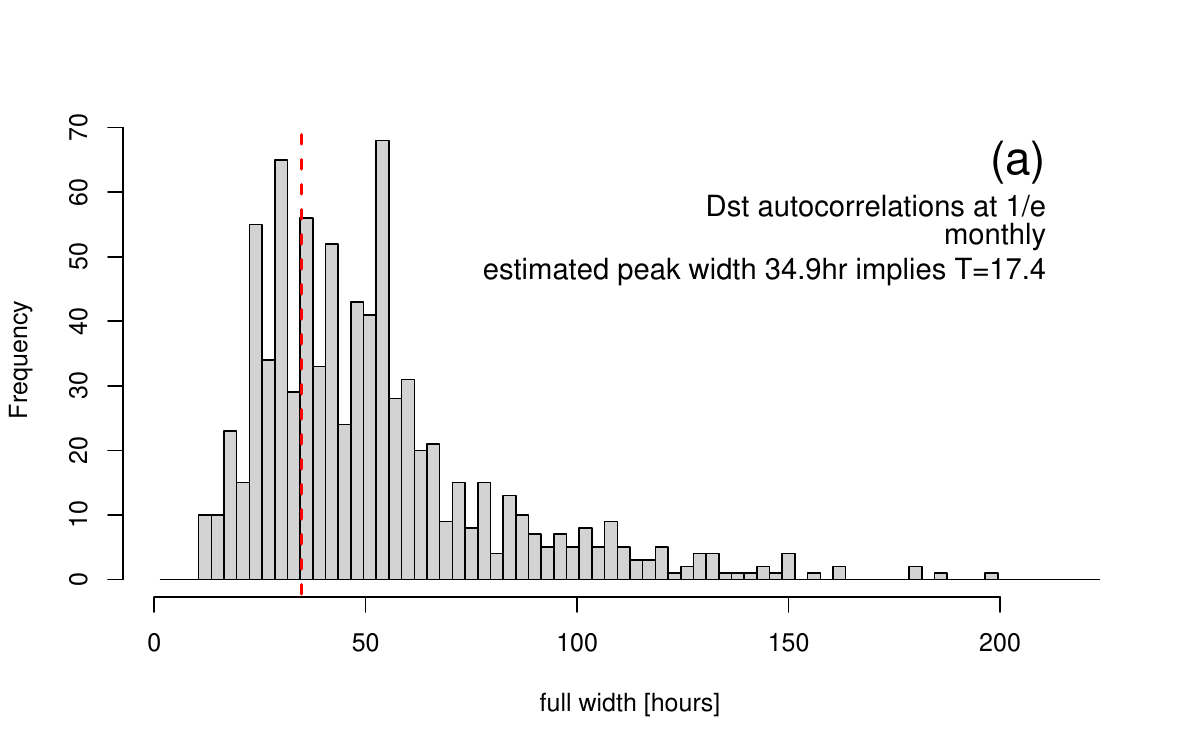}
\includegraphics[width=0.790\columnwidth]{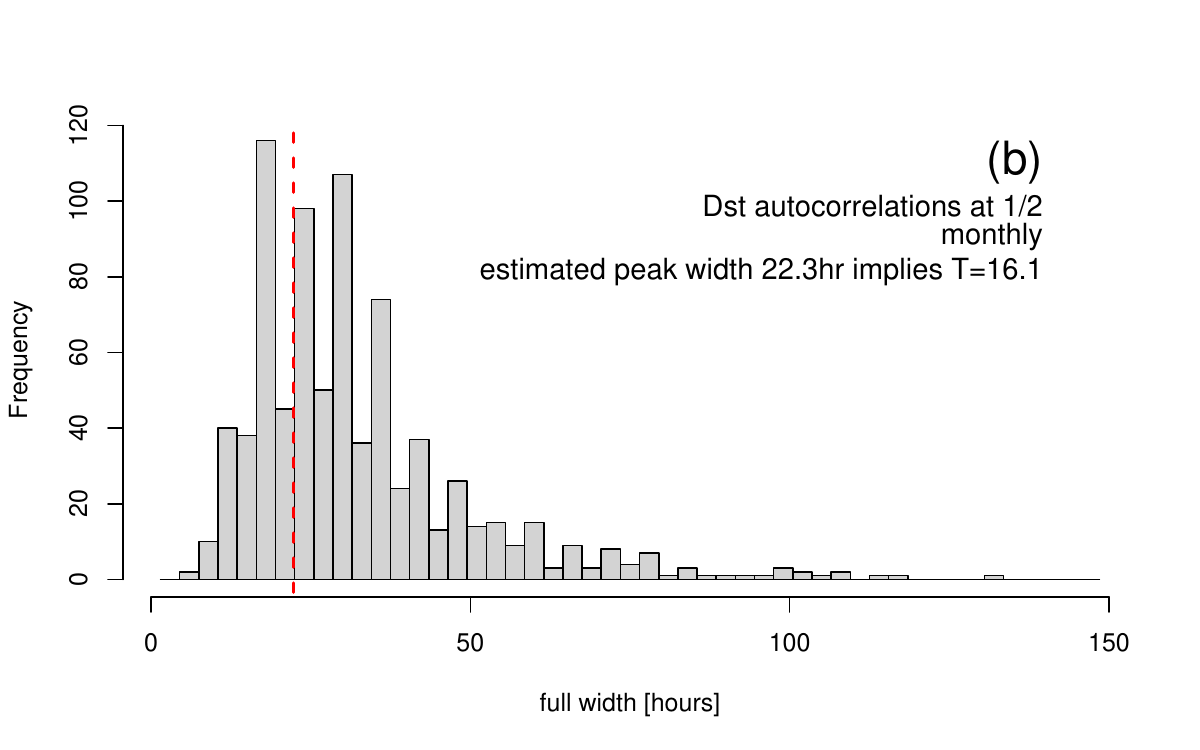}
\includegraphics[width=0.790\columnwidth]{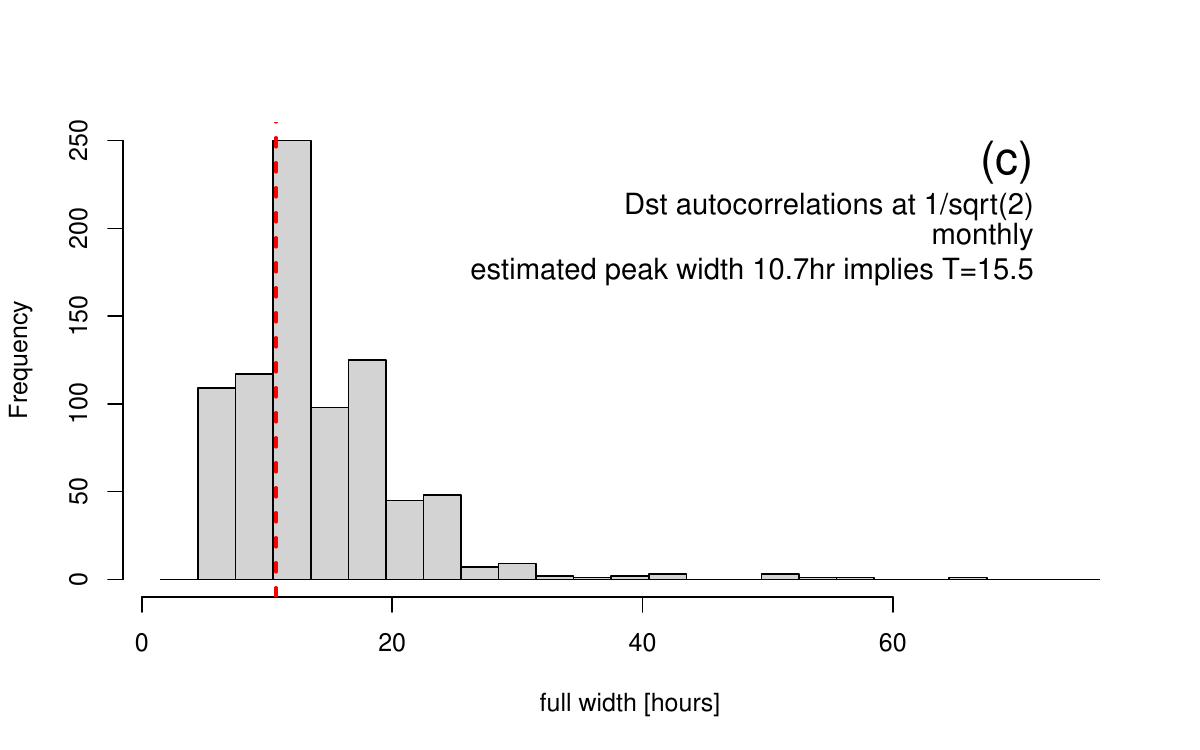}
\caption{Autocorrelation widths for the time series of ${\DS}$ values
 in monthly slices, 
 as binned into histograms.
It is to be expected that the higher value thresholds used when 
 calculating the autocorrelations widths
 gives rise to shorter times.
The red dashed lines indicate the peak position of a 
 LN fit to each histogram; 
 but note that case
 (c) suffers slightly from a lack of resolution
 due to its narrower range and the hourly cadence of ${\DS}$ data.
}
\label{fig-autocorr-histo}
\end{figure}

To implement this simple synthetic $\DS$ time series model 
 we need to estimate the timescales over which 
 subsequent ${\DS}$ values retain or lose correlations.
Naturally, 
 since they may not have a simple form, 
 there will be some approximations involved.
We can see from 
 figs. \ref{fig-autocorr} and \ref{fig-autocorr-histo}
 that ${\DS}$ timescales are limited to roughly
 the order of (several) tens of hours, 
 the particular value depending on the threshold criterion chosen.

In the above, 
 the choice of decay rate $\gamma$ 
 means that an autocorrelation full width for $Z$ at the $1/e$ level 
 will be $2T$; 
 at $1/2$ it will be $2 T \log(2) \simeq 1.4 T$, 
 and at $1/\sqrt{2}$ it will be $T \log(2) \simeq 0.7 T$.
Since it is reasonable to expect the timescales for ${\DS}$
 to be similar
 \footnote{Numerical experiment 
 suggests LN correlation times are almost the same
 as for its generating gaussian process $Z$.
},
 by inspecting fig. \ref{fig-autocorr-histo} (a,b),
 we see that we might reasonably 
 (or at least not-unreasonably) 
 assign $T \sim 16 $hr.

\begin{figure}
\includegraphics[width=0.790\columnwidth]{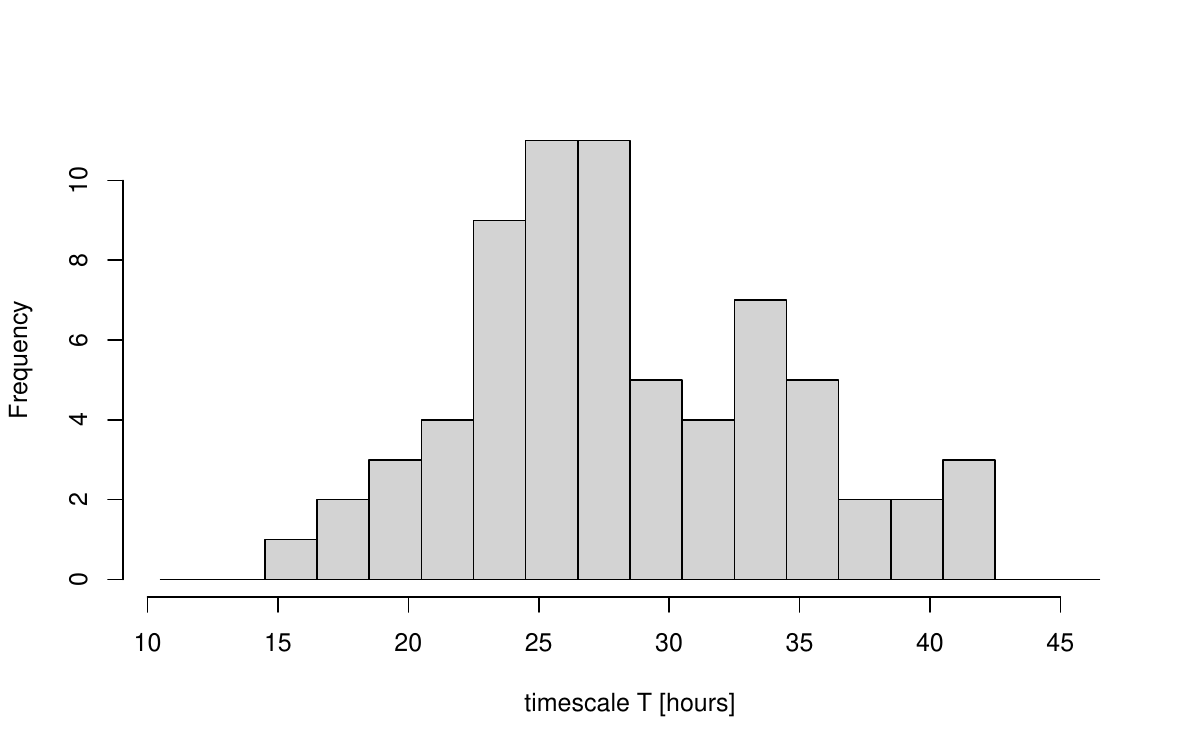}
\caption{Histogram of the best-matched $T$ values
 computed year-by-year
 over the whole 1957--2025 ${\KP}$ dataset.
}
\label{fig-Trefitting}
\end{figure}

As a caveat, 
 we should note that on fig. \ref{fig-autocorr-histo}
 the month-by-month autocorrelations widths
 show some considerable variation,
 meaning that the simple fixed-$\gamma$ stochastic process 
 of \eqref{eqn-dstcorrelates}
 is limited in its ability to truly represent a long-term
 ${\DS}$ time series.
However,
 as a check
 if we have a specific set of data (e.g. some particular year) in mind
 we can go further than working from the all-years distribution 
 of correlation times.
Given some specific long-enough ${\DS}$ dataset, 
 we can fit it as described above, 
 then use its offset ${\doff}$ to get the matching ${\DS}'$ sequence, 
 and take the logarithm to get a $Z$-like sequence.
Since this gives what we can hope will be a sequence that follows 
 \eqref{eqn-Zdiffusion}, 
 we can rearrange for the noise term $W(t)$
~
\begin{align}
  \sigma W(t)
&= 
  Z'(+1hr)
 -
  \mu
 -
  \frac{
    \left( Z(t) - \mu \right)
    \left( 1 - \gamma \right)
       }
  { \sqrt{ \left( 2\gamma \right)} }
.
\end{align}
If our model is accurate, 
 and in particular
 if $\gamma$ (or equivalently $T$) is correct,
 this inferred $W$ should be delta-correlated.
Therefore 
 what we can do is check this over a 
 range of $\gamma$  values
 in order to minimise the non-delta part
 of the autocorrelation function $\alpha(t-t')=\left<W(t)W(t')\right>$.
Using the sum of squares as a metric, 
 $\alpha(t-t')$ typically decreases sharply as $T$
 increases from zero towards ten hours, 
 but then flattens and rises again --
 albeit only very slightly; 
 indicating the distinct insensitivity 
 of this test to any specific long-enough choice of $T$.
Fig. \ref{fig-Trefitting}
 shows a histogram of the best-matched $T$ values
 for the whole 1957--2025 ${\DS}$ dataset;
 note that the peak best-matched $T$ value
 is about half-again larger 
 than those on fig. \ref{fig-autocorr-histo}(a,b).
This could be due to either or 
 (or perhaps both) 
 or two factors: 
 a greater role for longer correlations, 
 and/or 
 the imperfect match between the LN and the 
 actual ${\DS}$ distributions.

Here I have tested the performance based on the calendar year fits
 that are the main focus of this report.
However, 
 any sufficiently long time series of ${\DS}$
 could be used to construct fitting parameters; 
 thus if interested in making a sequence  
 for some particular date, 
 one could simply fit the immediately preceeding year's worth 
 of ${\DS}$ values, 
 and use those parameters to generate the time series.

%
\section{Summary}\label{S-summary}

Here I demonstrated that the log-normal distribution 
 could be used to fit year-long ${\DS}$ distributions.
Although an analysis of the Kullback-Leibler divergence 
 between those fits and the data showed 
 an imperfect statistical match, 
 the fits nevertheless provide a good visual match 
 and good estimates of basic statistical properties. 
Since ${\DS}$ values 
 lie on a continuum, 
 the matches were achieved by allowing for a per-year customised offset 
 between raw values and those fitted.
Further, 
 autocorrelation times for ${\DS}$ time series' were considered, 
 and a simple model for generating
 synthetic ${\DS}$ sequences proposed.

%
%

\clearpage
\appendix
\section{Year-by-year fits}\label{S-yearfits}

\begin{figure}
\includegraphics[width=0.30\columnwidth]{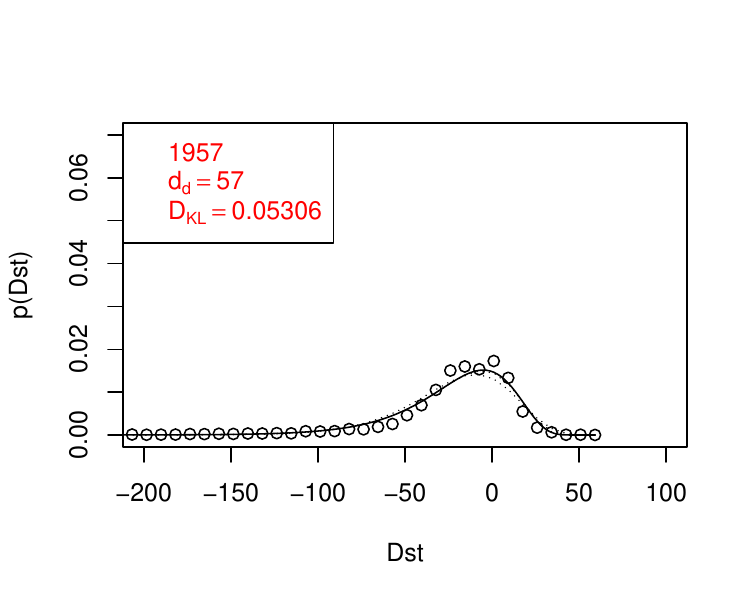}
\includegraphics[width=0.30\columnwidth]{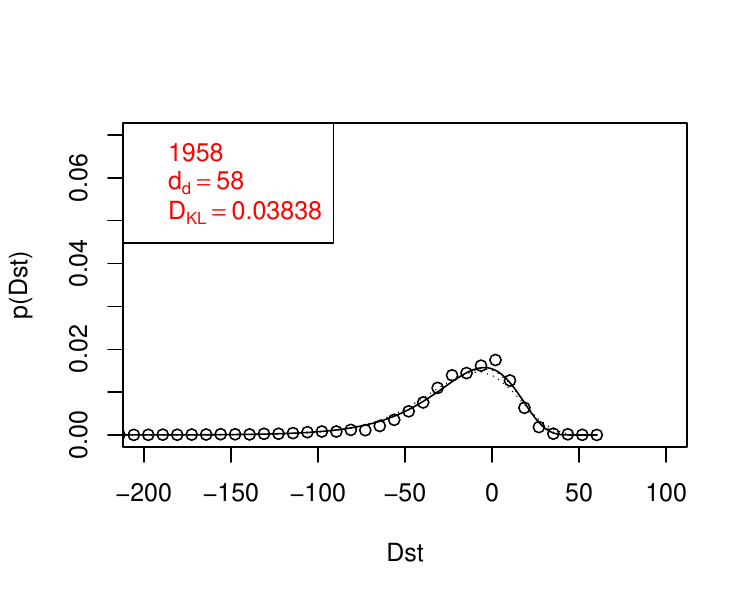}
\includegraphics[width=0.30\columnwidth]{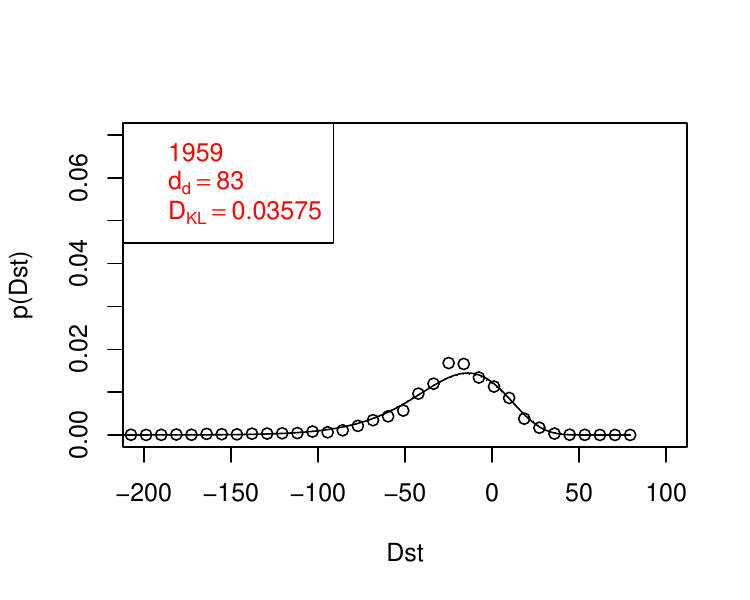}
\includegraphics[width=0.30\columnwidth]{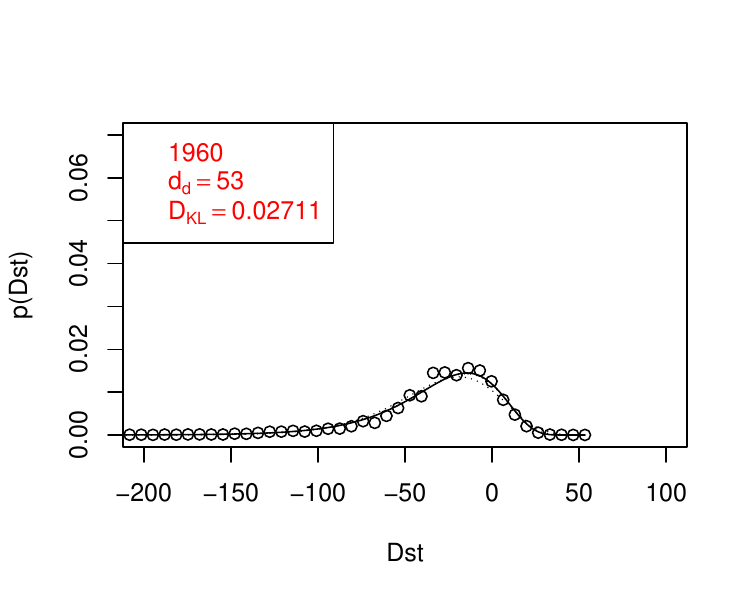}
\includegraphics[width=0.30\columnwidth]{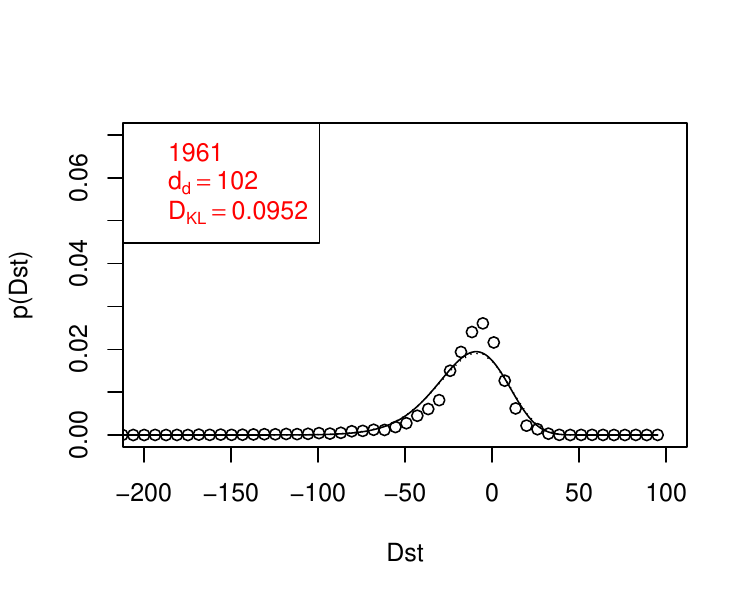}
\includegraphics[width=0.30\columnwidth]{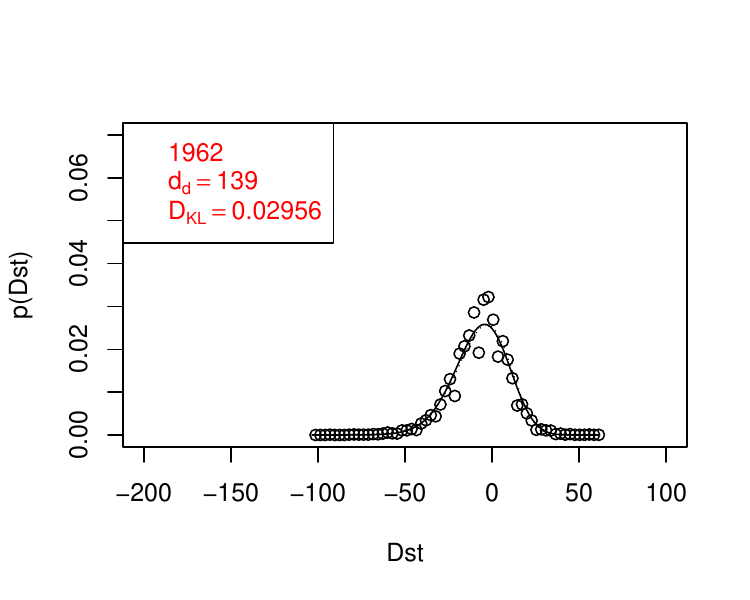}
\includegraphics[width=0.30\columnwidth]{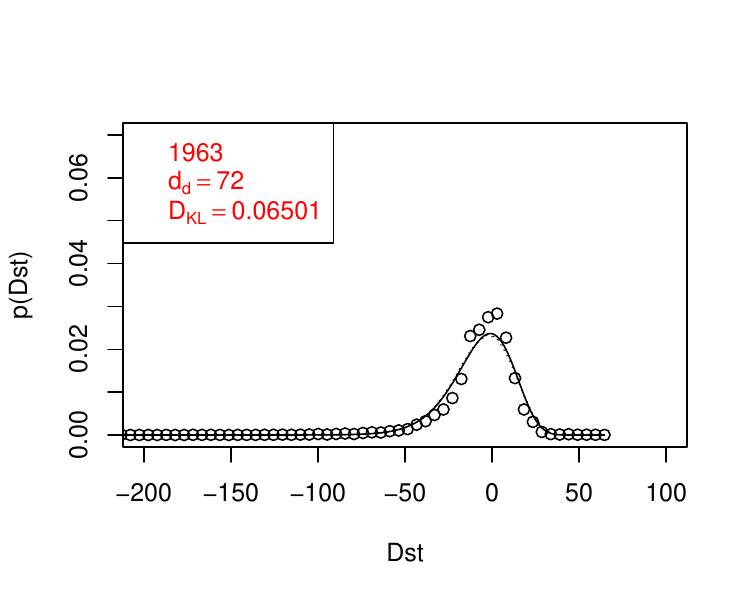}
\caption{Solar cycle 19, i.e. 1954--1963.}
\label{fig-scycle19}
\end{figure}
\begin{figure}
\includegraphics[width=0.30\columnwidth]{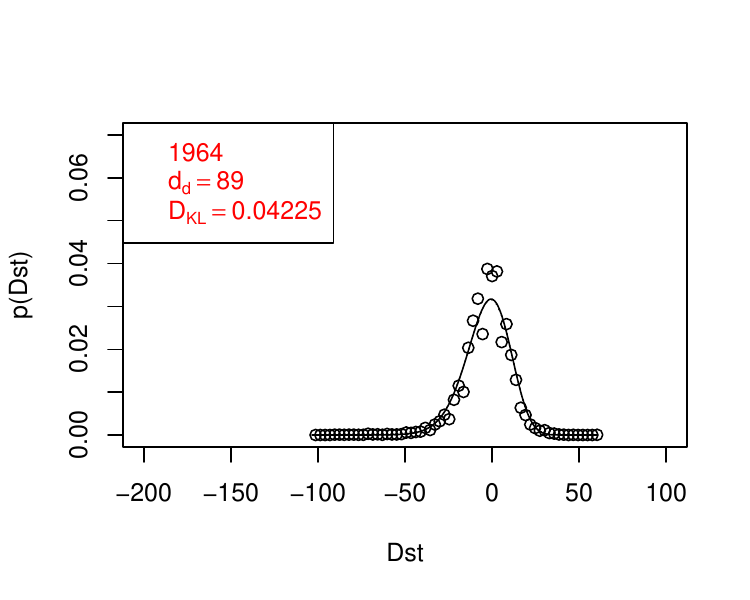}
\includegraphics[width=0.30\columnwidth]{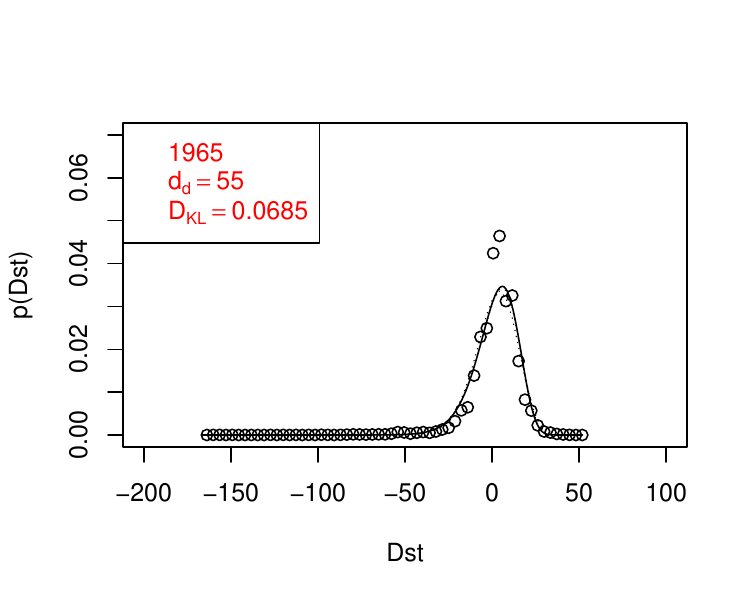}
\includegraphics[width=0.30\columnwidth]{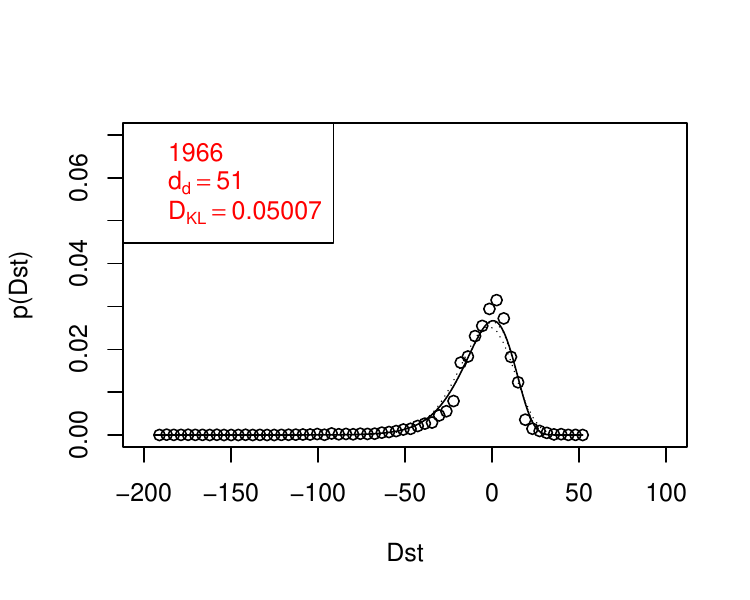}
\includegraphics[width=0.30\columnwidth]{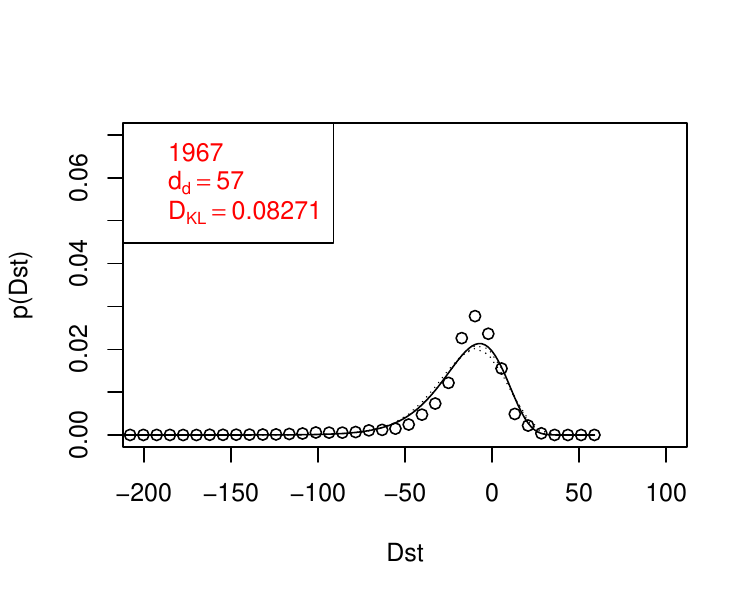}
\includegraphics[width=0.30\columnwidth]{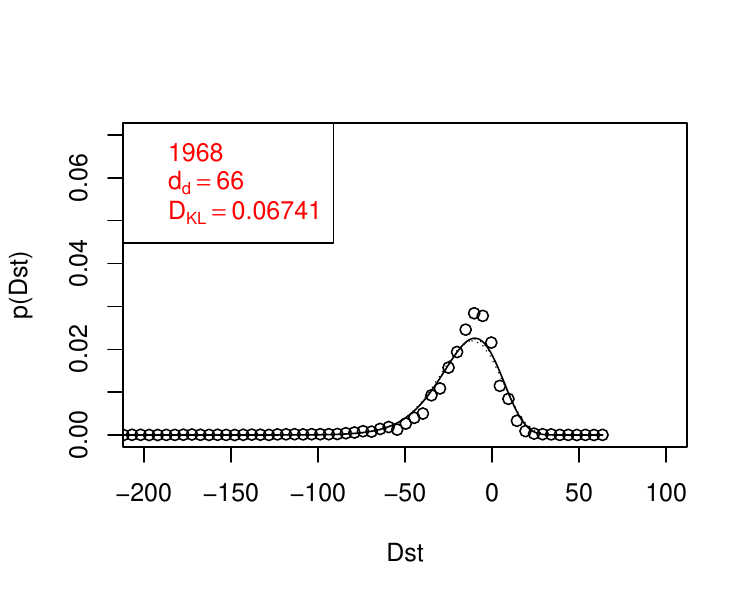}
\includegraphics[width=0.30\columnwidth]{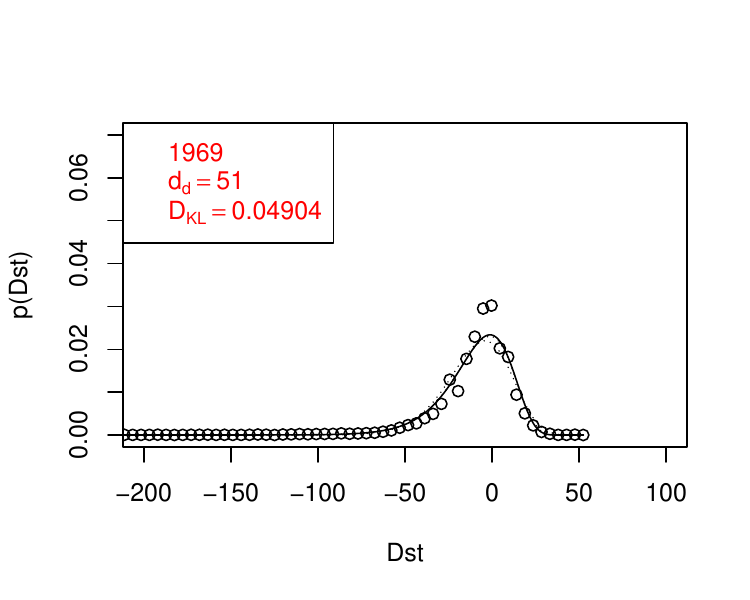}
\includegraphics[width=0.30\columnwidth]{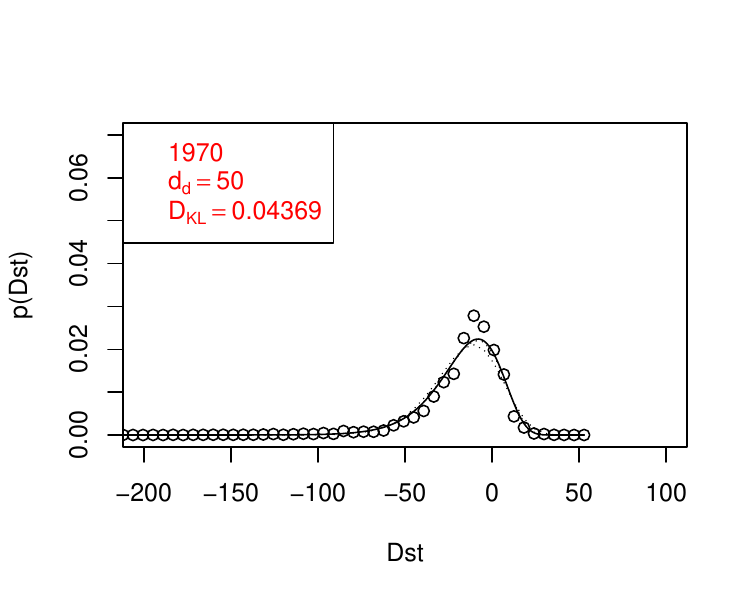}
\includegraphics[width=0.30\columnwidth]{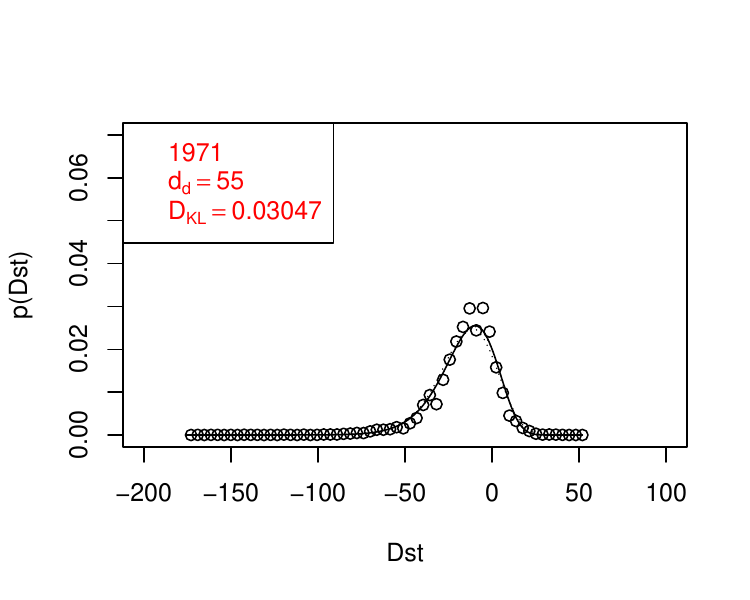}
\includegraphics[width=0.30\columnwidth]{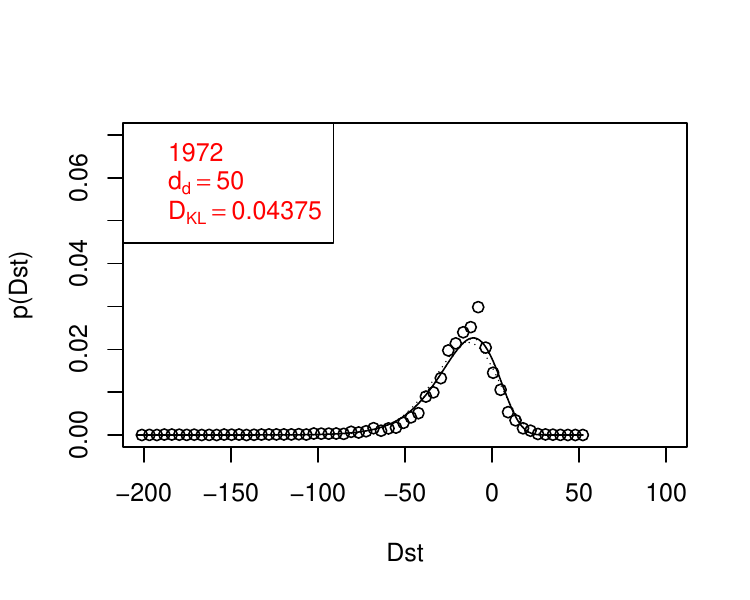}
\includegraphics[width=0.30\columnwidth]{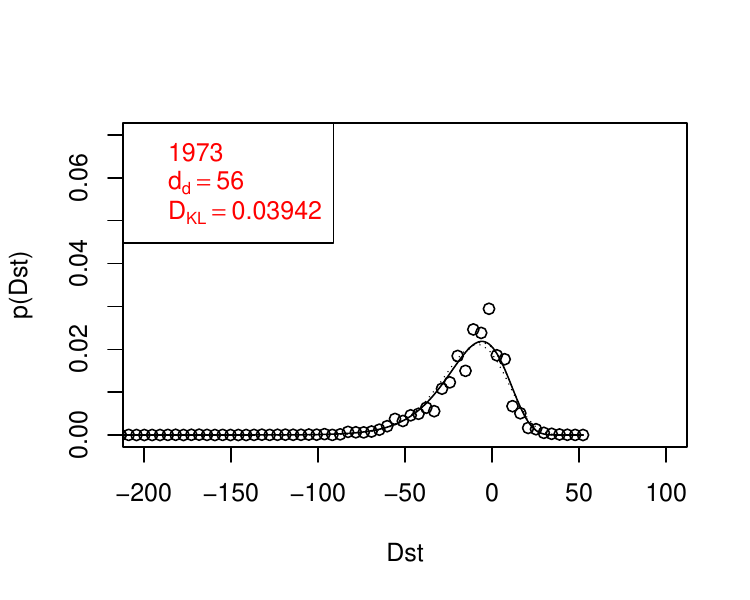}
\includegraphics[width=0.30\columnwidth]{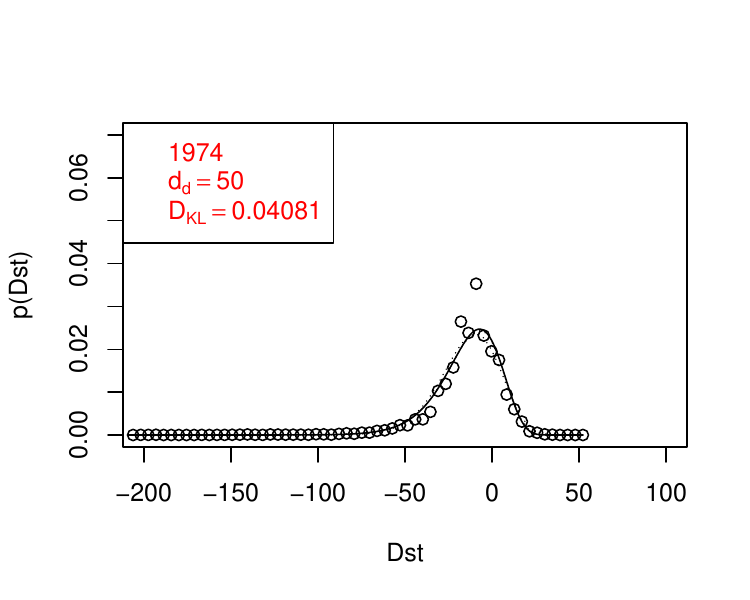}
\includegraphics[width=0.30\columnwidth]{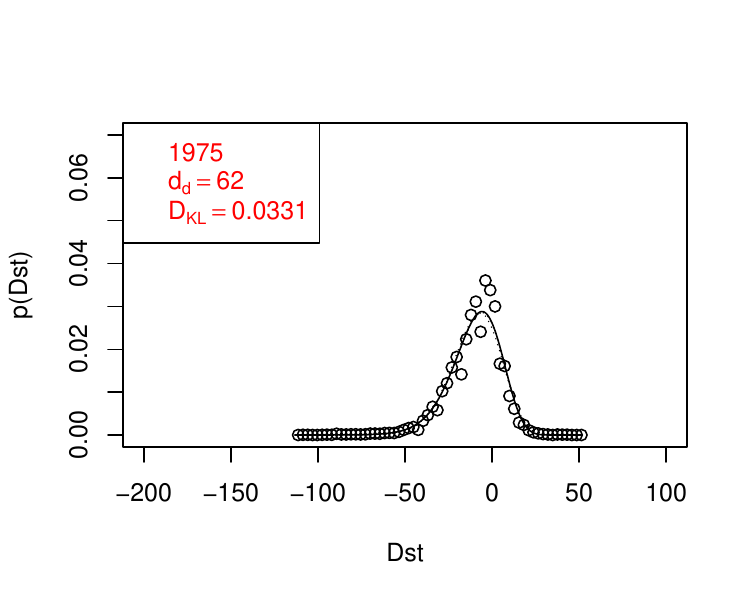}
\caption{Solar cycle 20, i.e. 1964--1975.}
\label{fig-scycle20}
\end{figure}
\begin{figure}
\includegraphics[width=0.30\columnwidth]{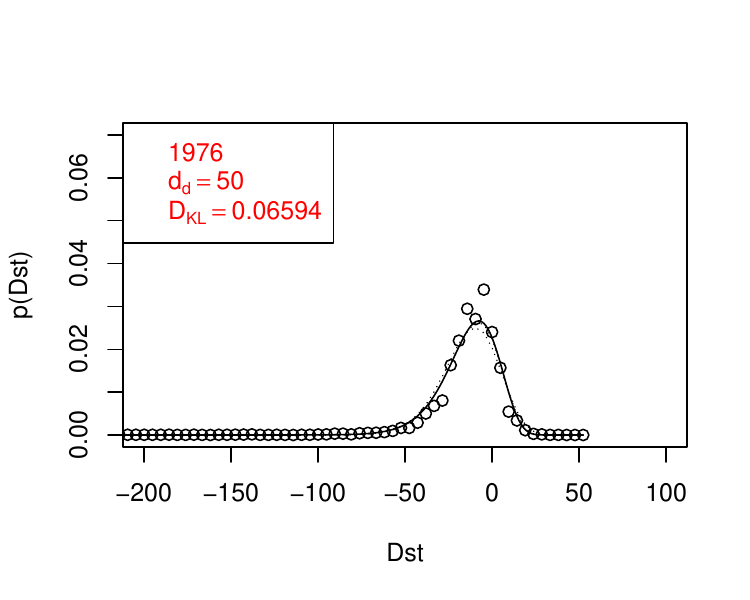}
\includegraphics[width=0.30\columnwidth]{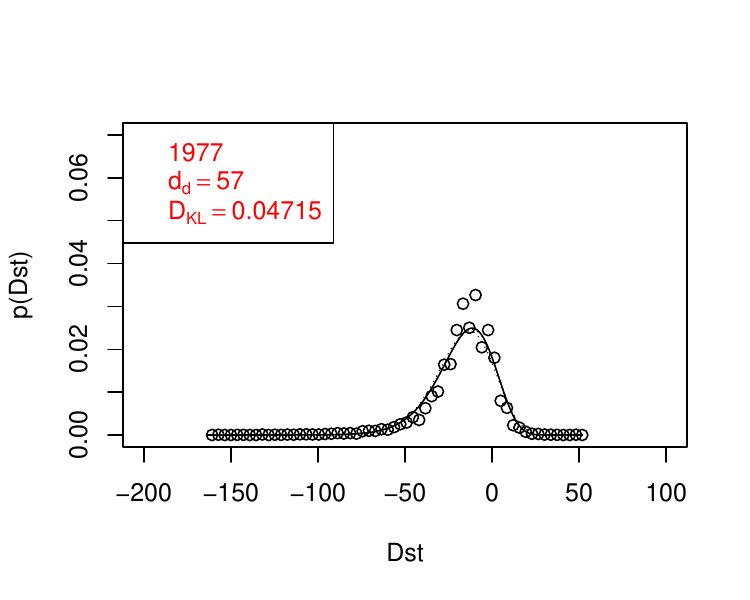}
\includegraphics[width=0.30\columnwidth]{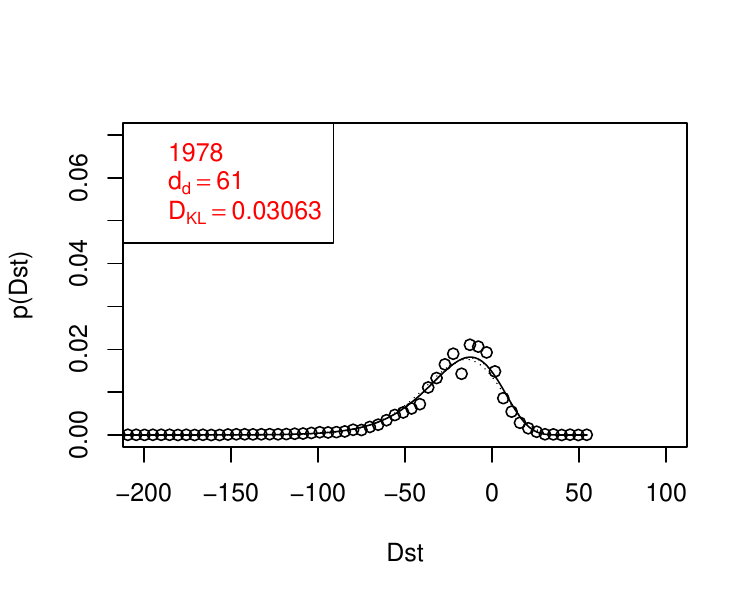}
\includegraphics[width=0.30\columnwidth]{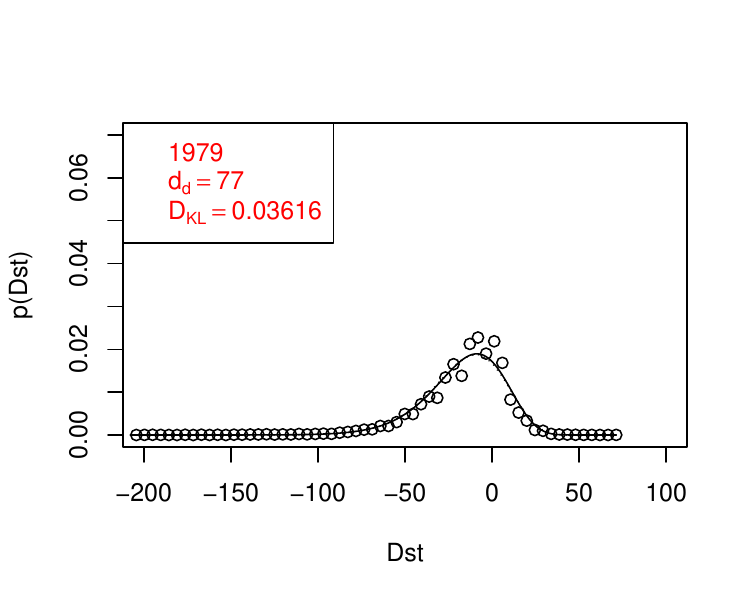}
\includegraphics[width=0.30\columnwidth]{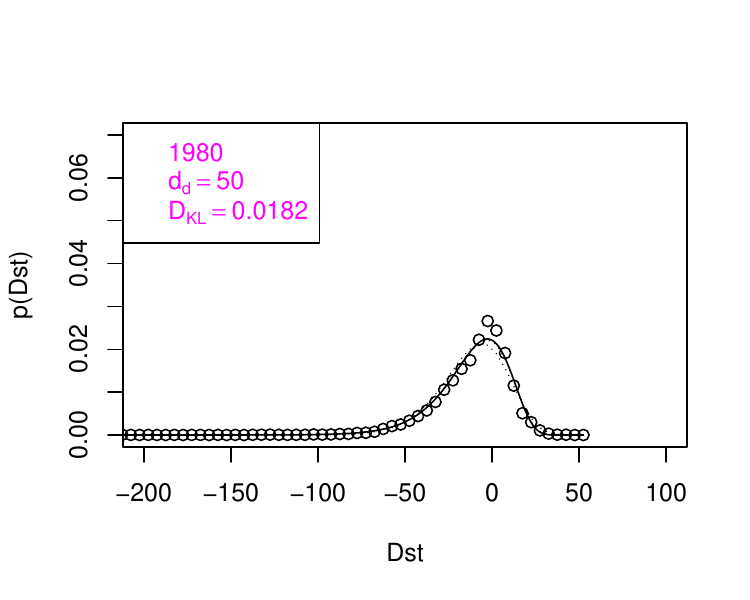}
\includegraphics[width=0.30\columnwidth]{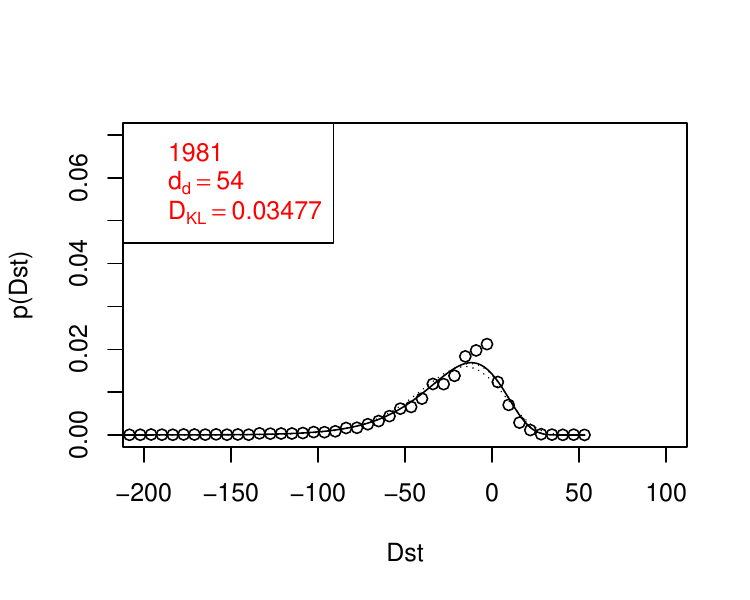}
\includegraphics[width=0.30\columnwidth]{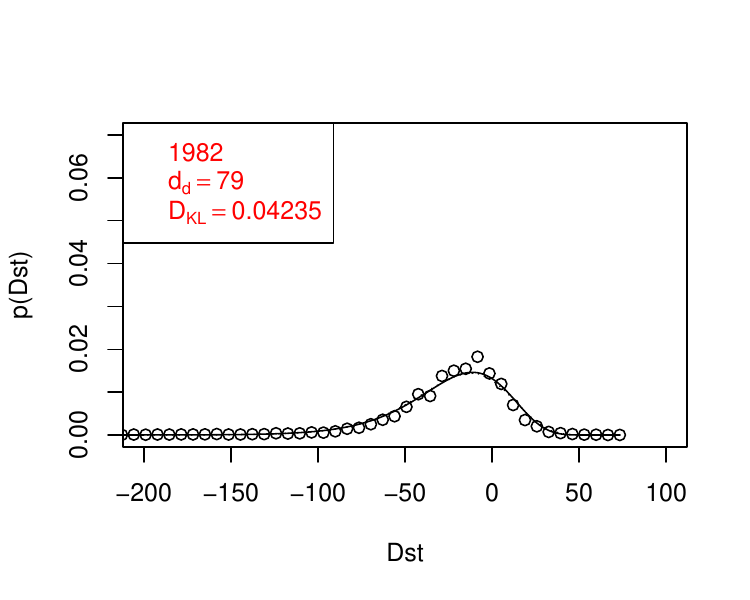}
\includegraphics[width=0.30\columnwidth]{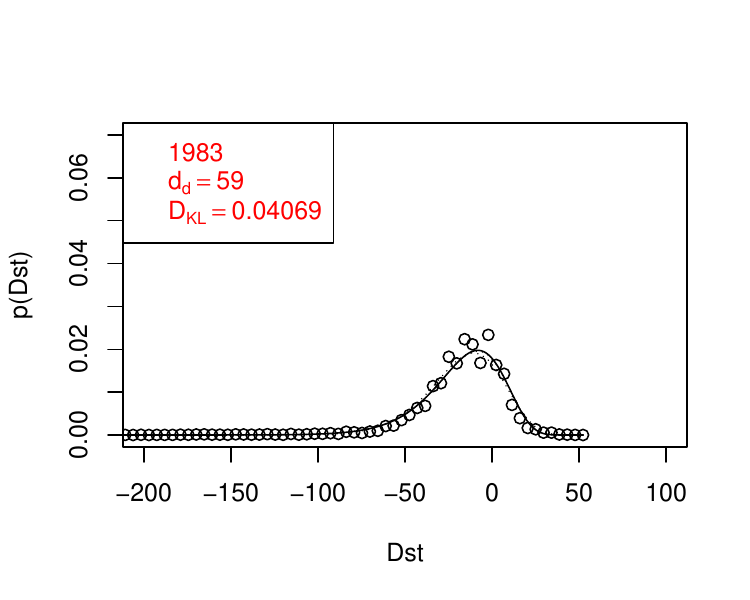}
\includegraphics[width=0.30\columnwidth]{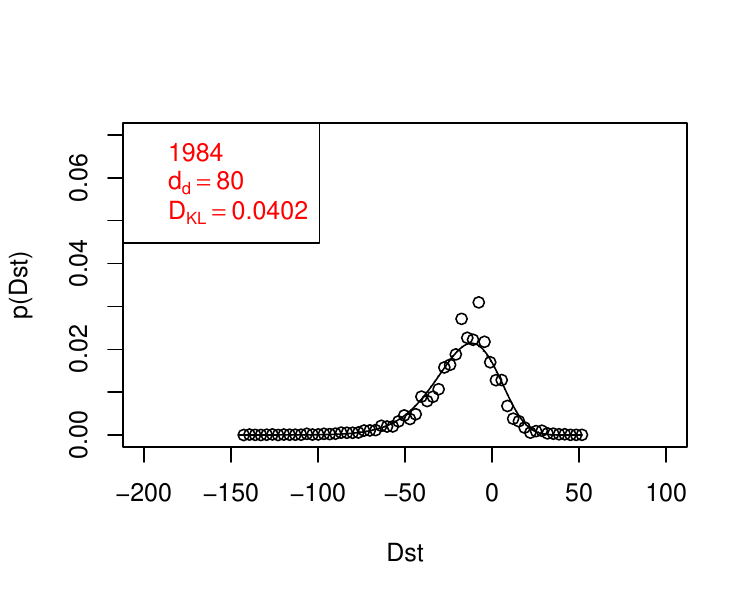}
\includegraphics[width=0.30\columnwidth]{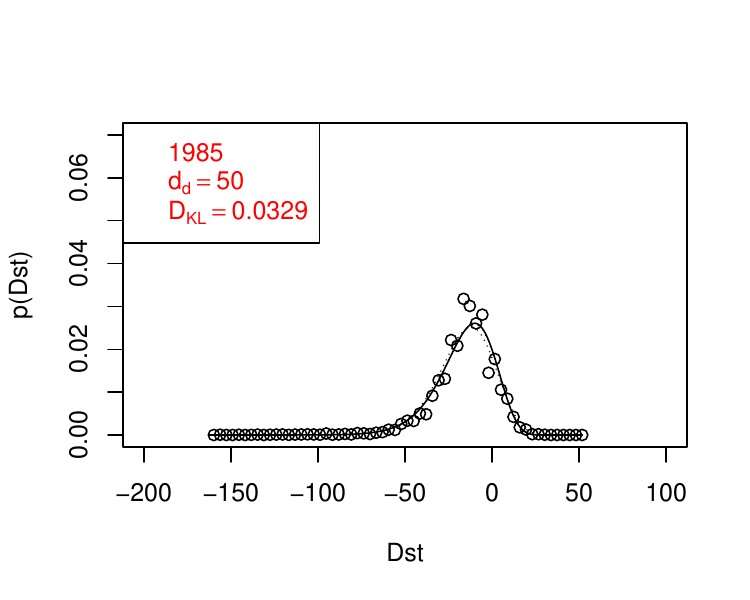}
\caption{Solar cycle 21, i.e. 1976--1985.}
\label{fig-scycle21}
\end{figure}
\begin{figure}
\includegraphics[width=0.30\columnwidth]{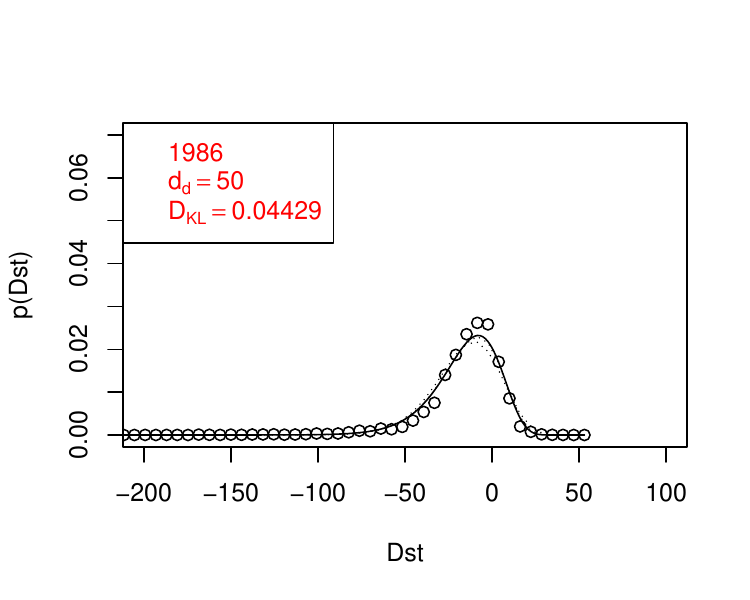}
\includegraphics[width=0.30\columnwidth]{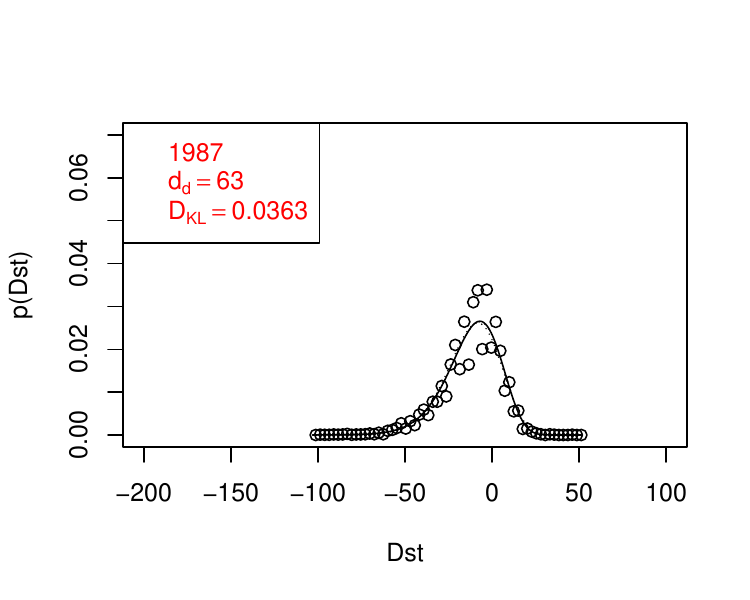}
\includegraphics[width=0.30\columnwidth]{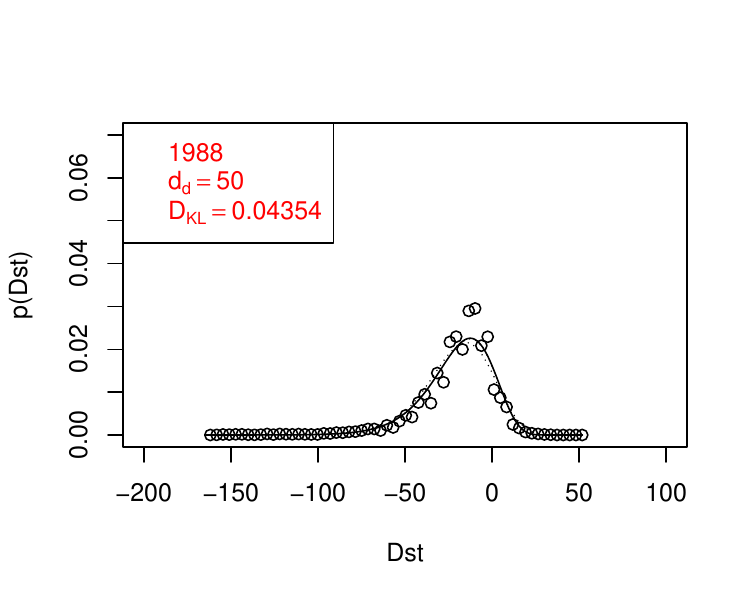}
\includegraphics[width=0.30\columnwidth]{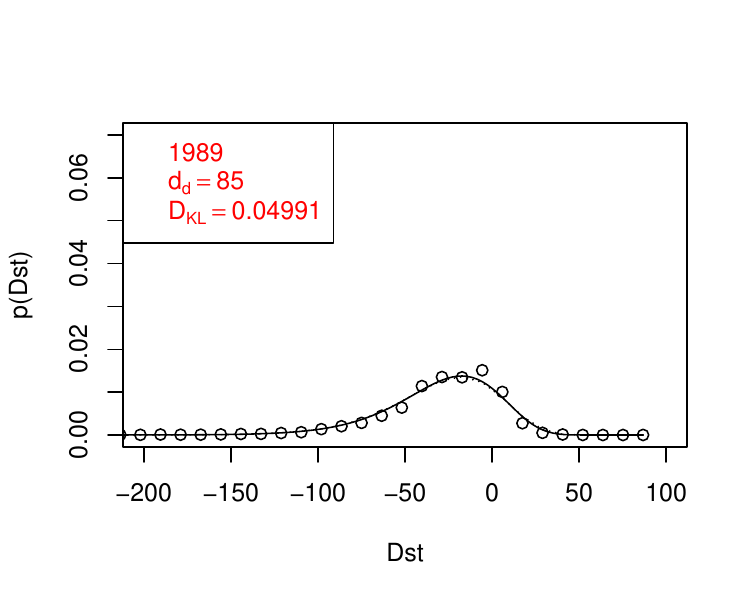}
\includegraphics[width=0.30\columnwidth]{Figs/figdat1990.lst.pdf}
\includegraphics[width=0.30\columnwidth]{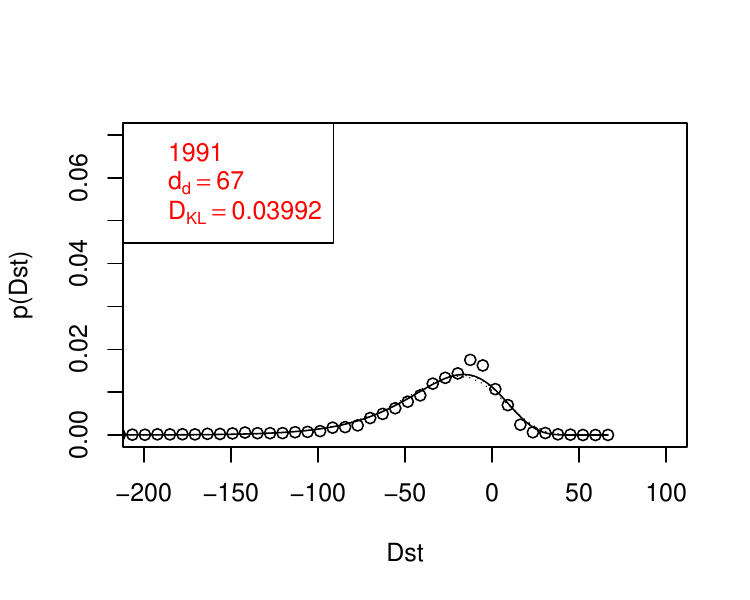}
\includegraphics[width=0.30\columnwidth]{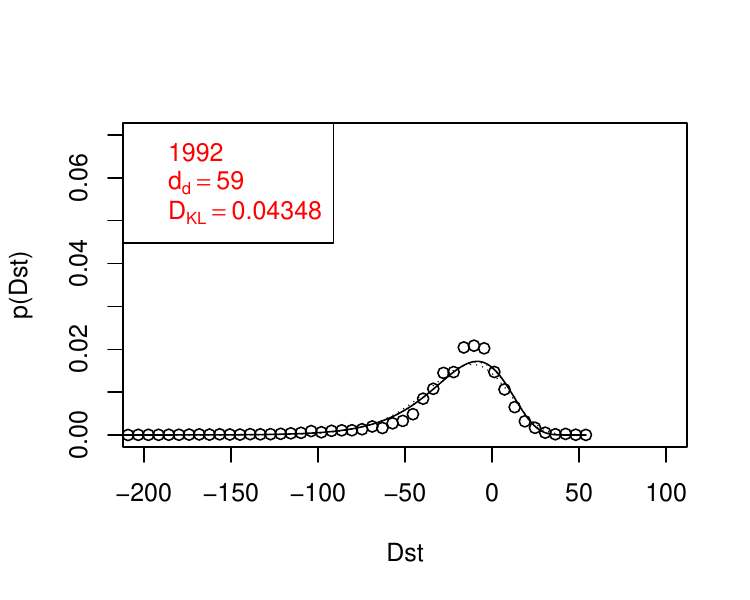}
\includegraphics[width=0.30\columnwidth]{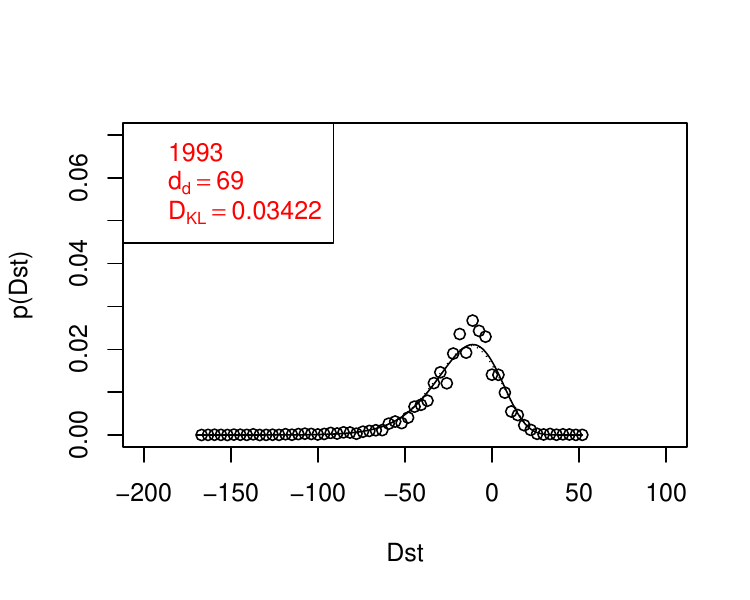}
\includegraphics[width=0.30\columnwidth]{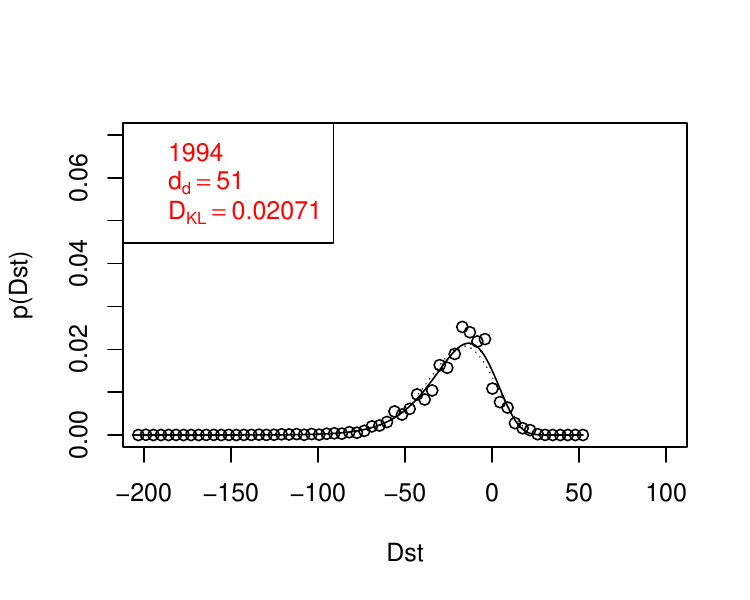}
\includegraphics[width=0.30\columnwidth]{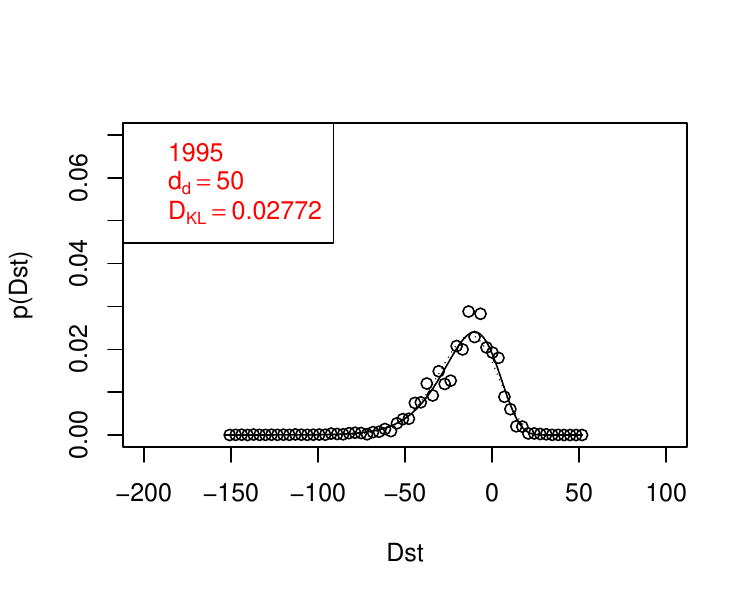}
\caption{Solar cycle 22, i.e. 1986-1995.}
\label{fig-scycle22}
\end{figure}
\begin{figure}
\includegraphics[width=0.30\columnwidth]{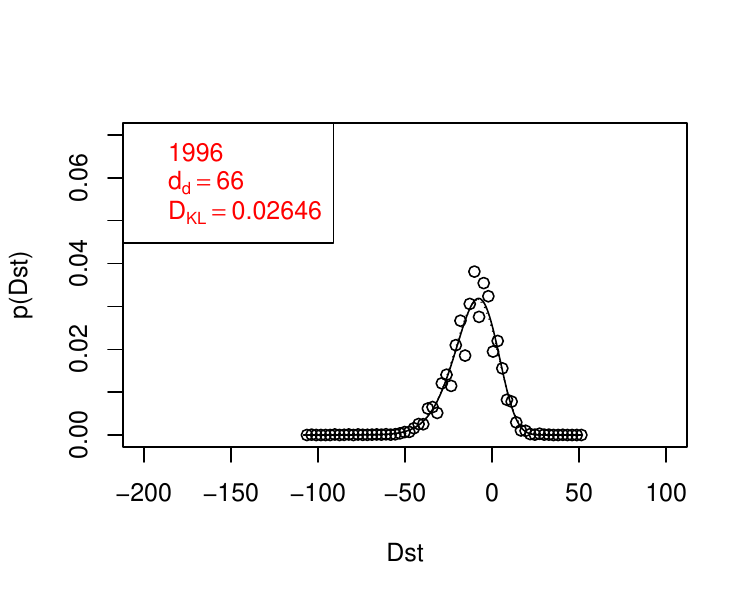}
\includegraphics[width=0.30\columnwidth]{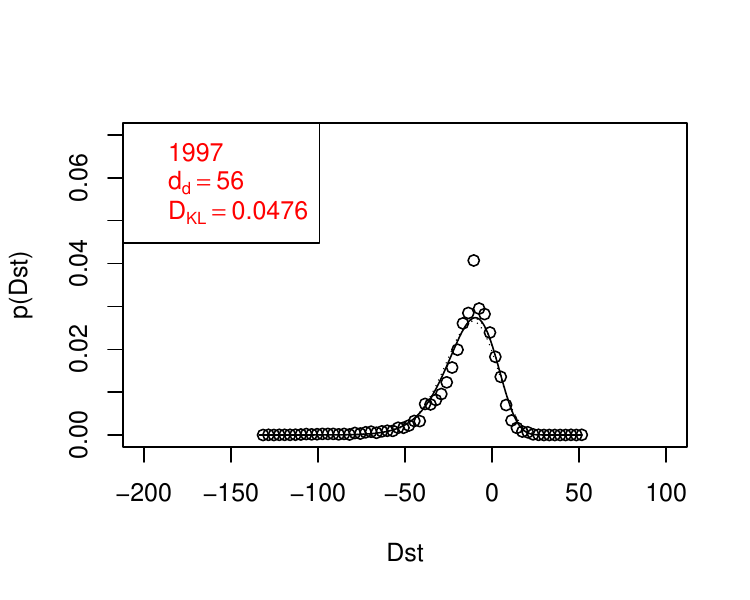}
\includegraphics[width=0.30\columnwidth]{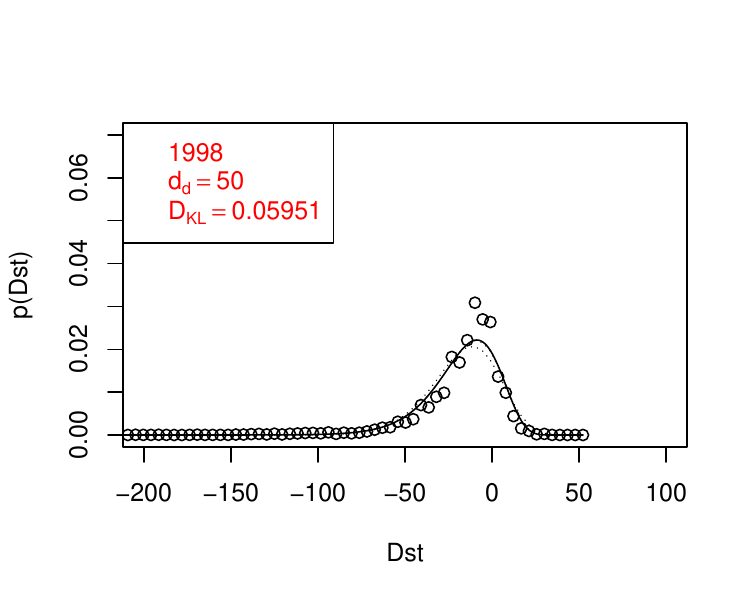}
\includegraphics[width=0.30\columnwidth]{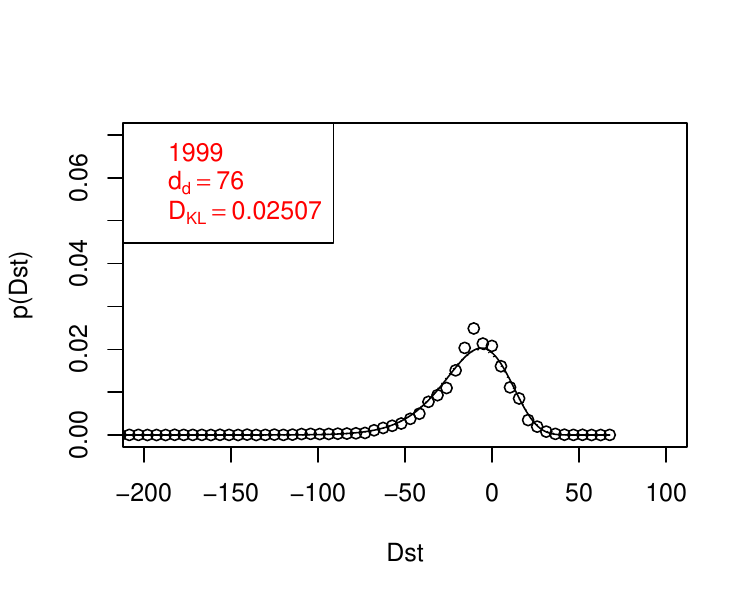}
\includegraphics[width=0.30\columnwidth]{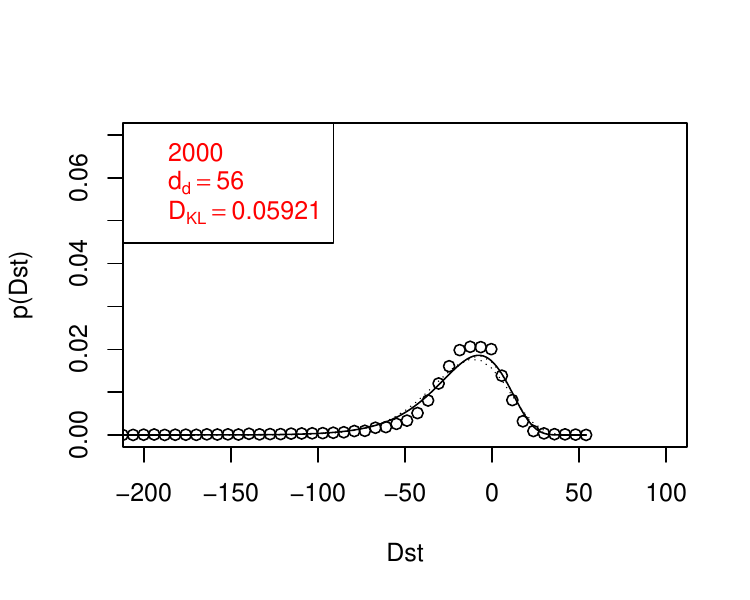}
\includegraphics[width=0.30\columnwidth]{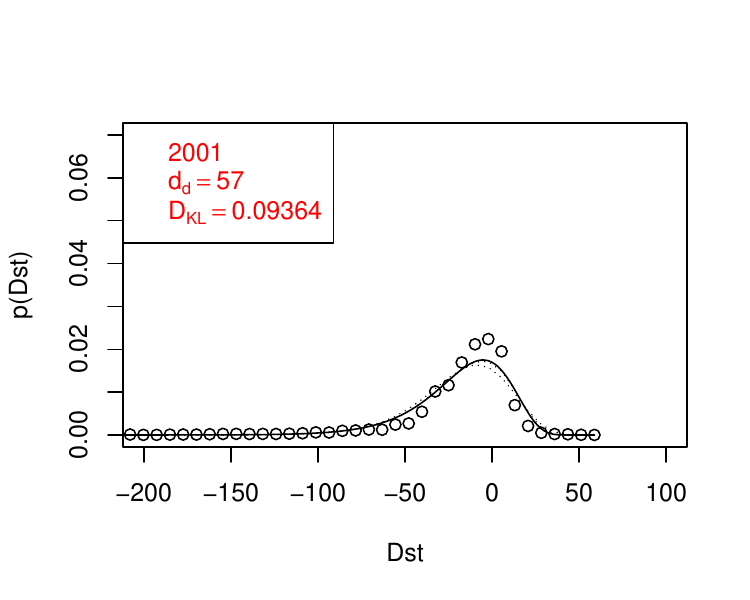}
\includegraphics[width=0.30\columnwidth]{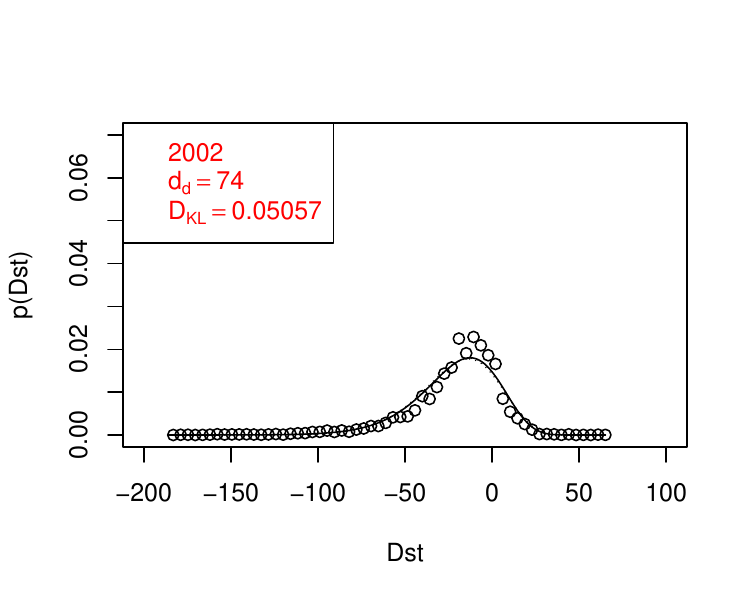}
\includegraphics[width=0.30\columnwidth]{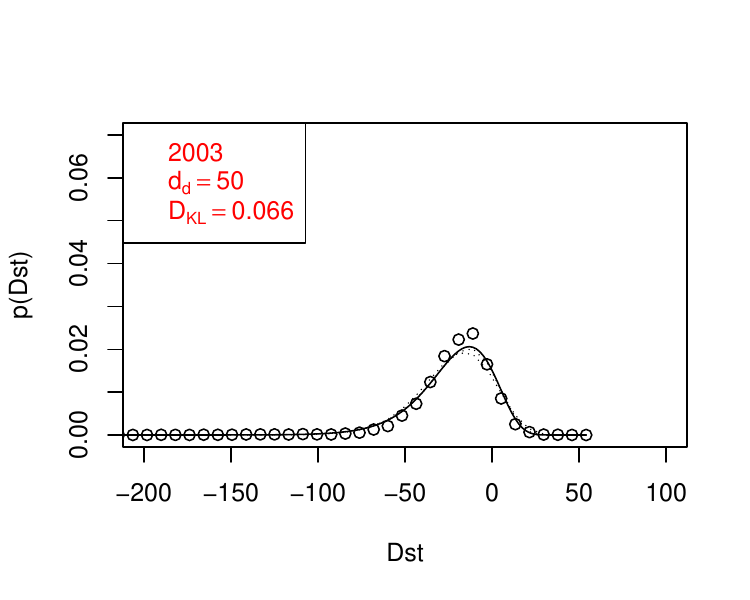}
\includegraphics[width=0.30\columnwidth]{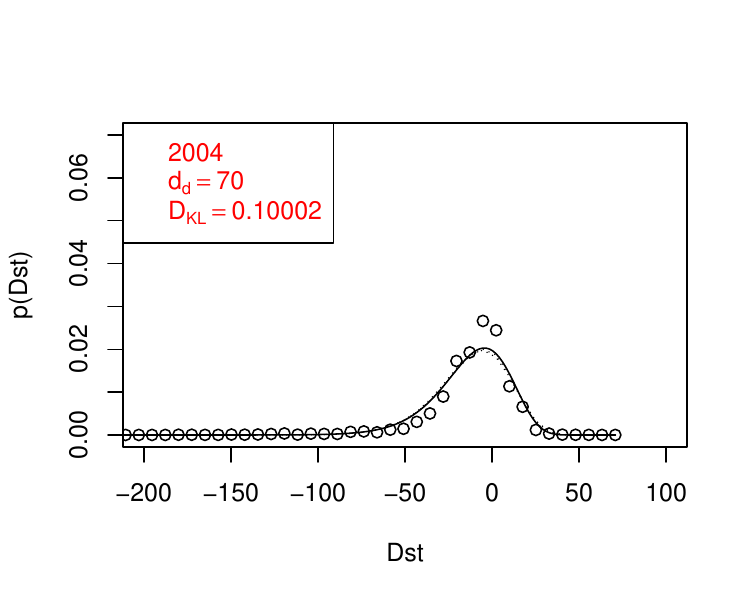}
\includegraphics[width=0.30\columnwidth]{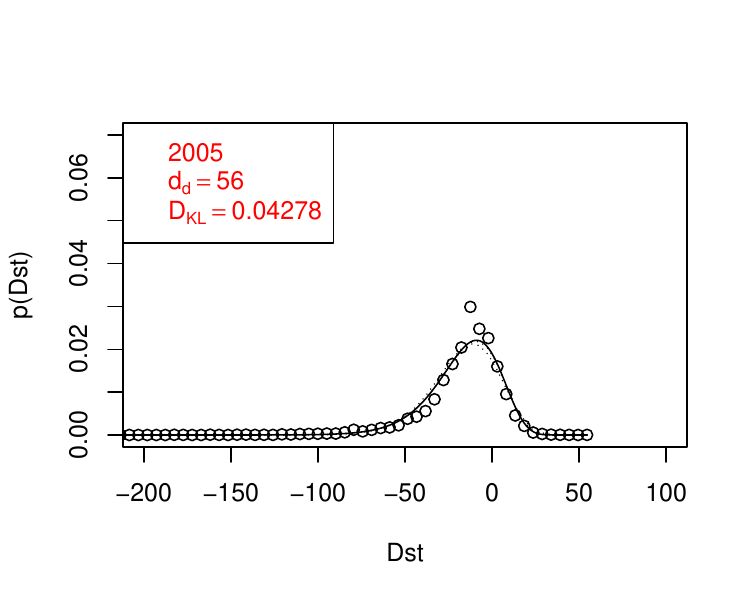}
\includegraphics[width=0.30\columnwidth]{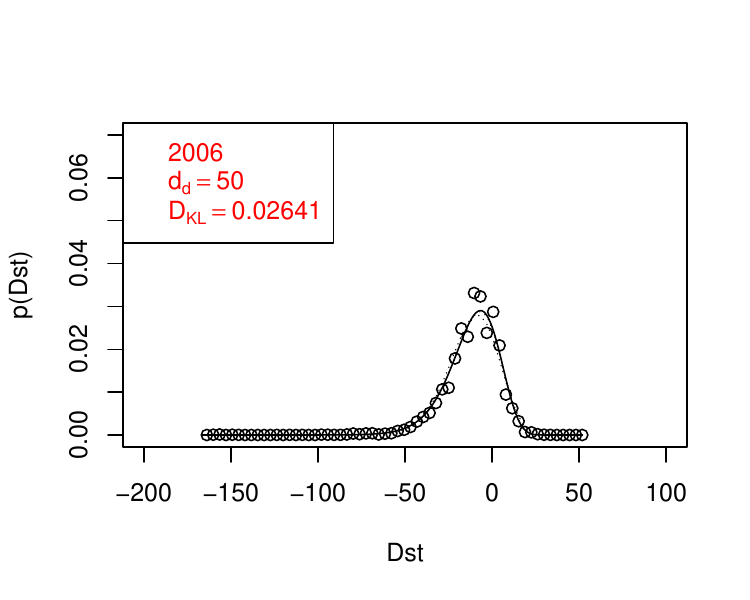}
\includegraphics[width=0.30\columnwidth]{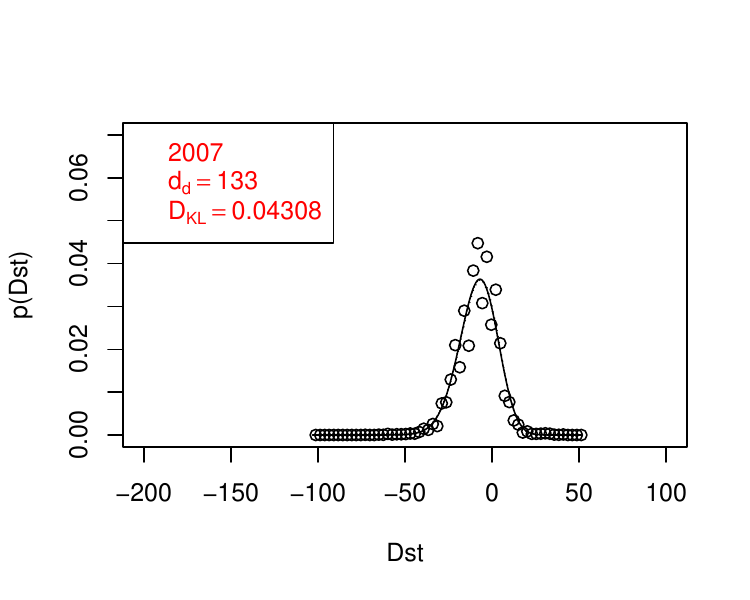}
\caption{Solar cycle 23, i.e. 1996--2007.}
\label{fig-scycle23}
\end{figure}
\begin{figure}
\includegraphics[width=0.30\columnwidth]{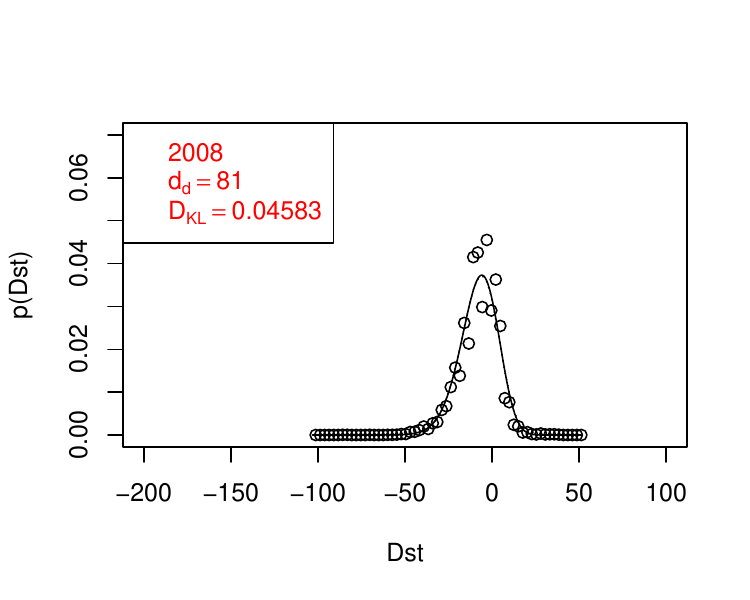}
\includegraphics[width=0.30\columnwidth]{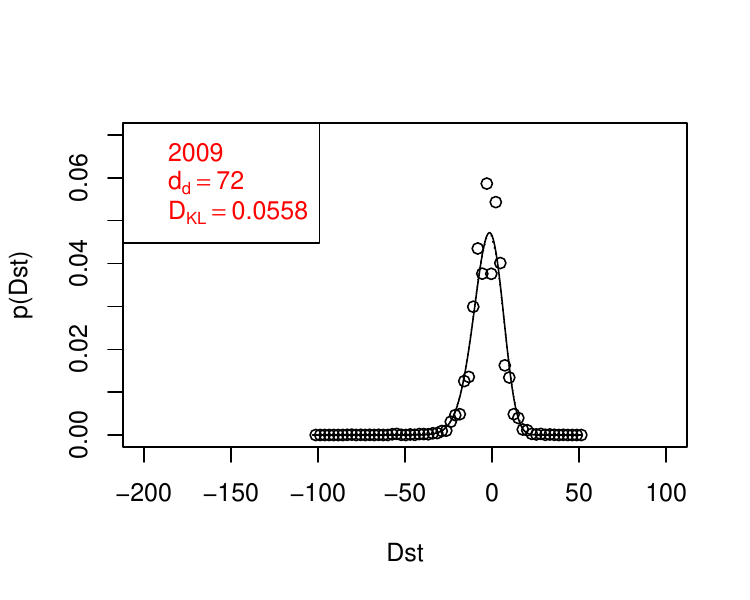}
\includegraphics[width=0.30\columnwidth]{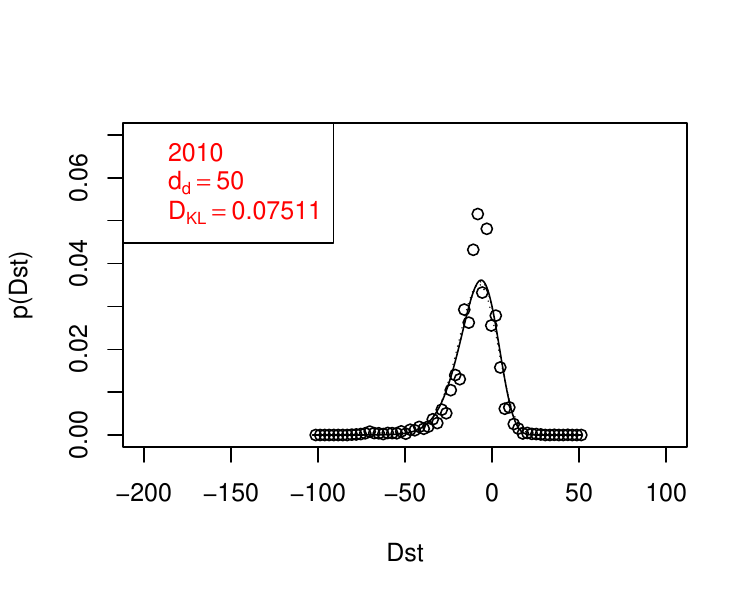}
\includegraphics[width=0.30\columnwidth]{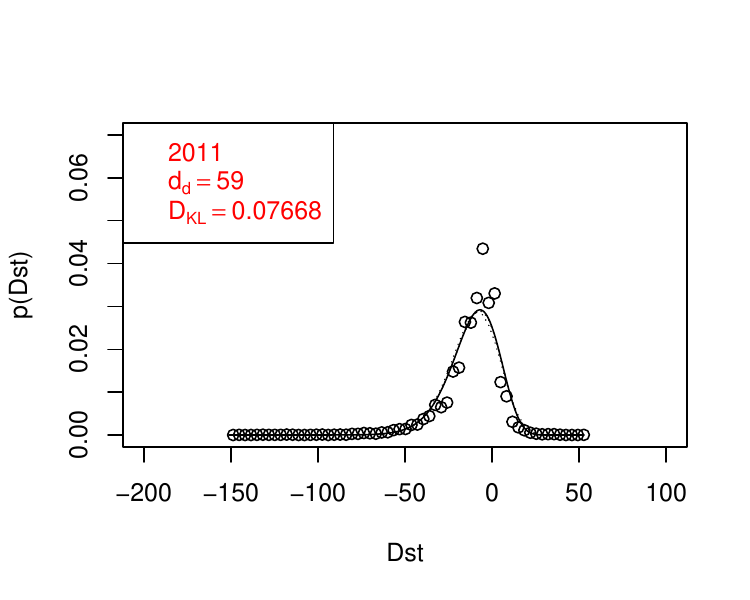}
\includegraphics[width=0.30\columnwidth]{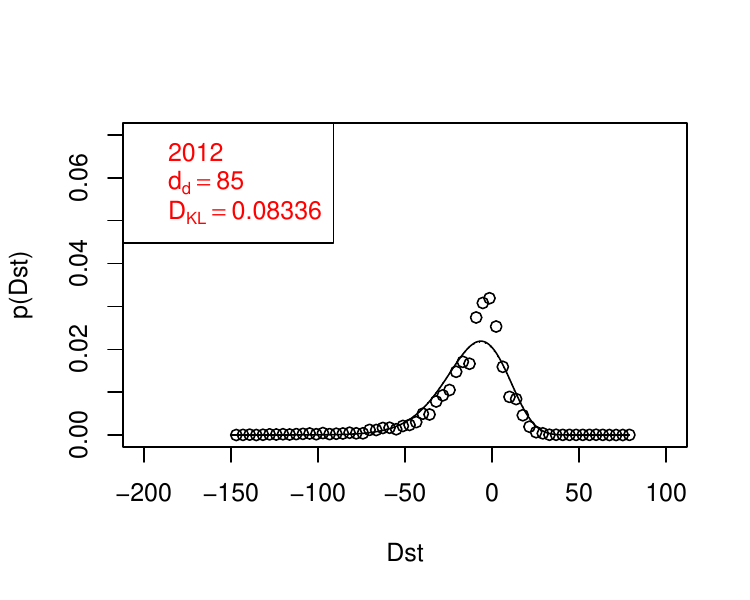}
\includegraphics[width=0.30\columnwidth]{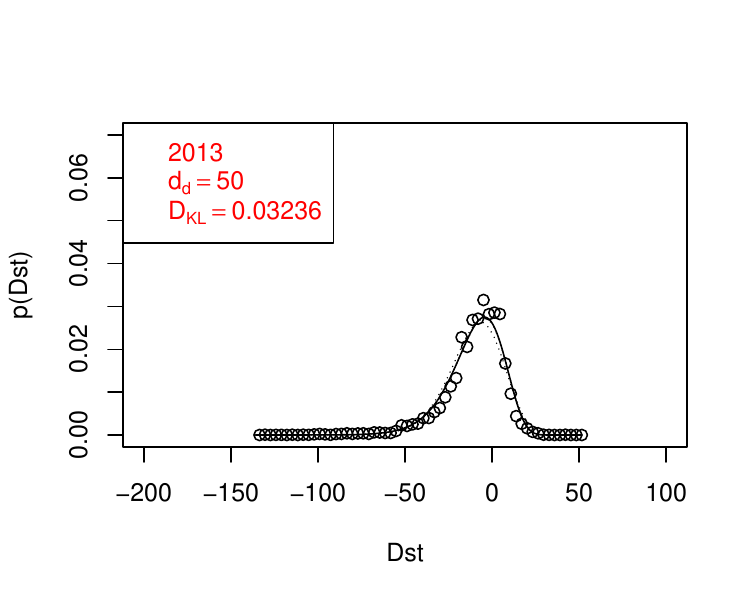}
\includegraphics[width=0.30\columnwidth]{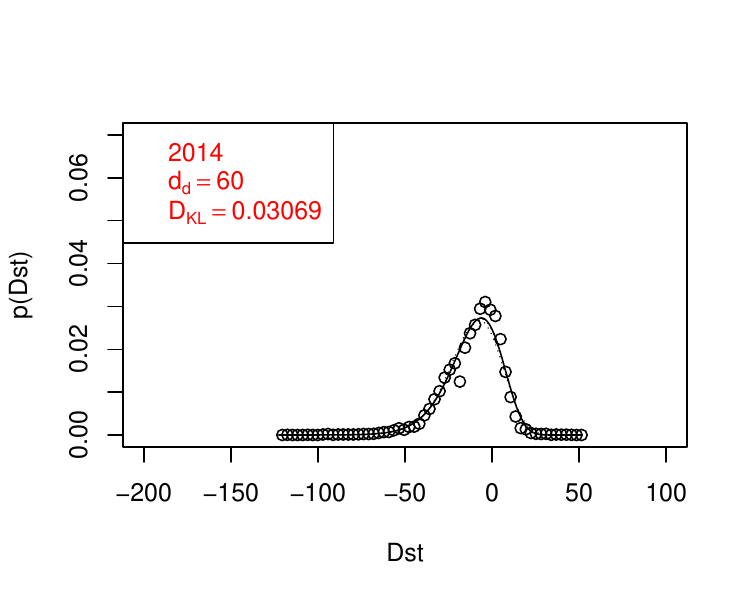}
\includegraphics[width=0.30\columnwidth]{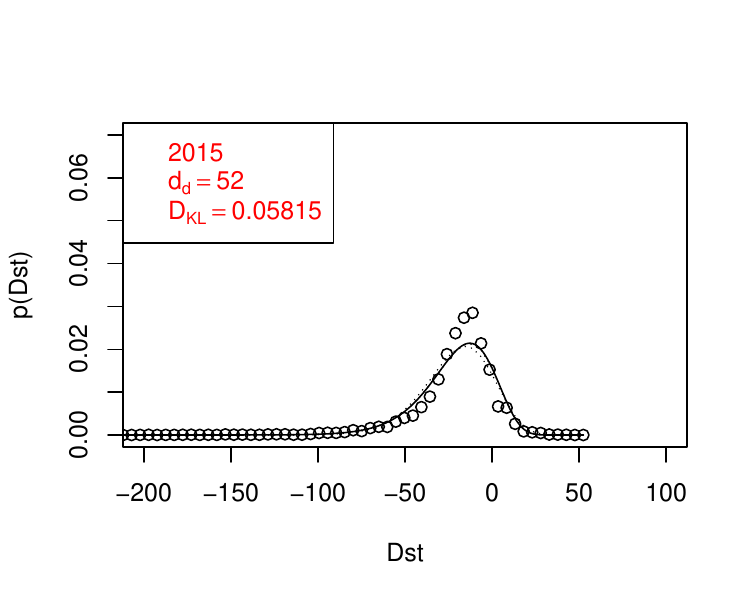}
\includegraphics[width=0.30\columnwidth]{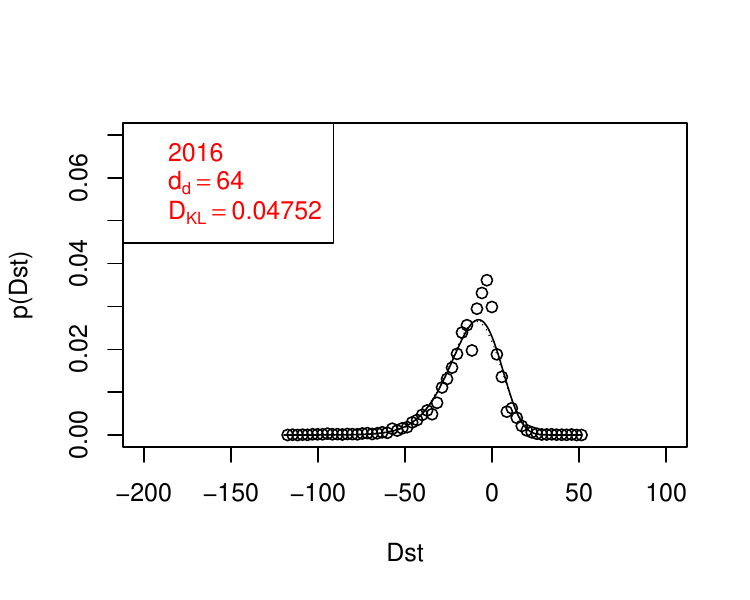}
\includegraphics[width=0.30\columnwidth]{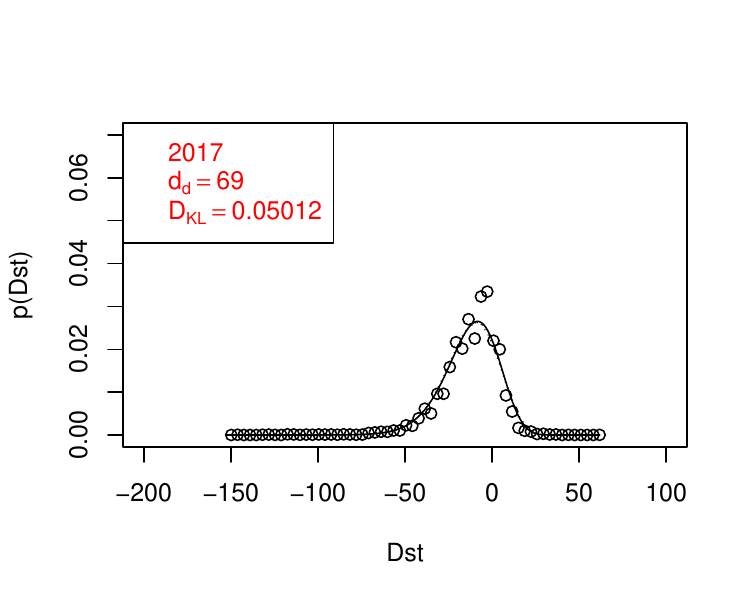}
\includegraphics[width=0.30\columnwidth]{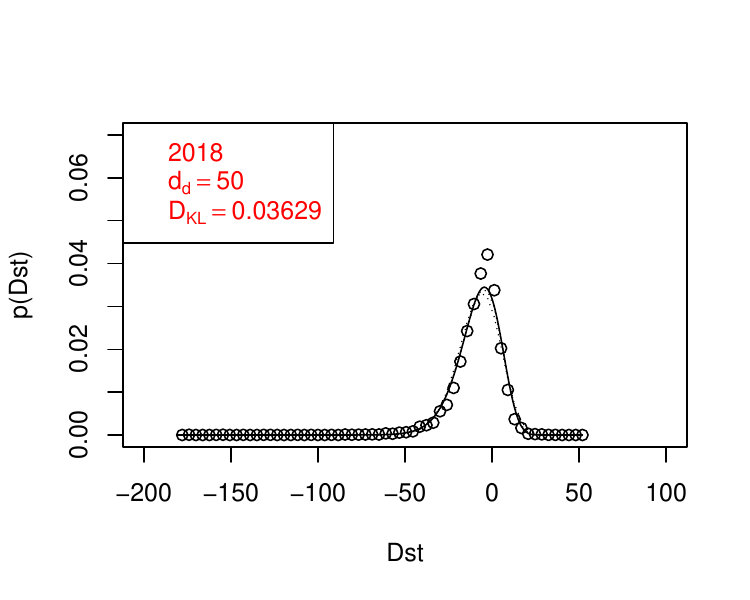}
\caption{Solar cycle 24, i.e. 2008--2018.}
\label{fig-scycle24}
\end{figure}
\begin{figure}
\includegraphics[width=0.30\columnwidth]{Figs/figdat2019.lst.pdf}
\includegraphics[width=0.30\columnwidth]{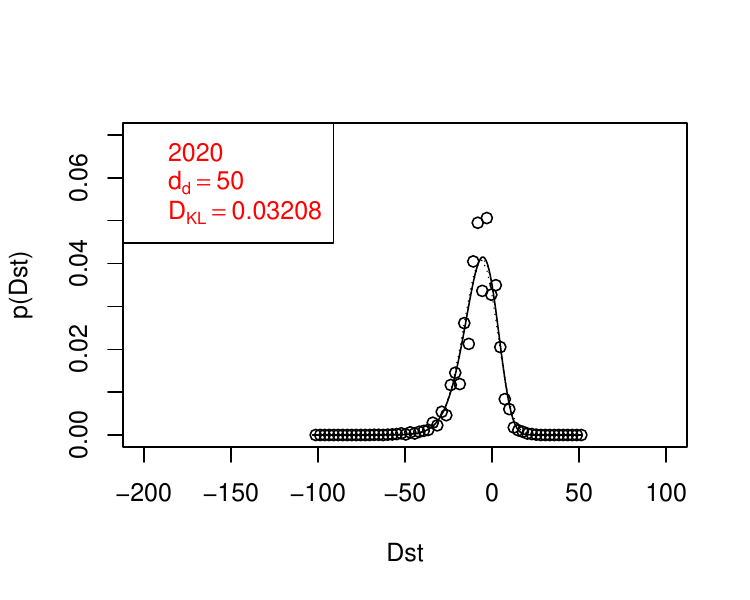}
\includegraphics[width=0.30\columnwidth]{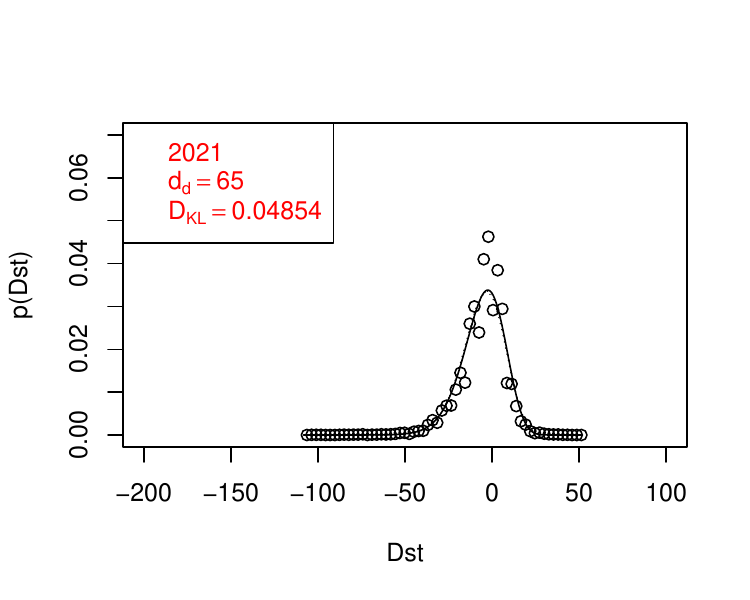}
\includegraphics[width=0.30\columnwidth]{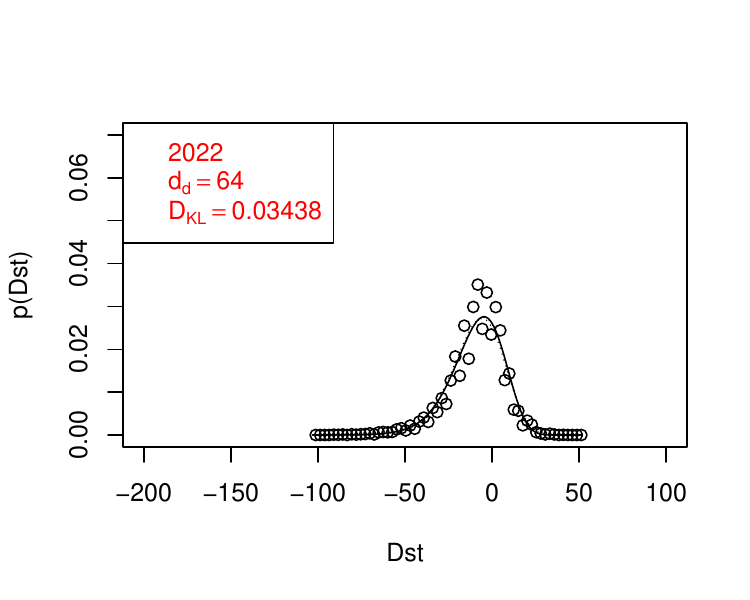}
\includegraphics[width=0.30\columnwidth]{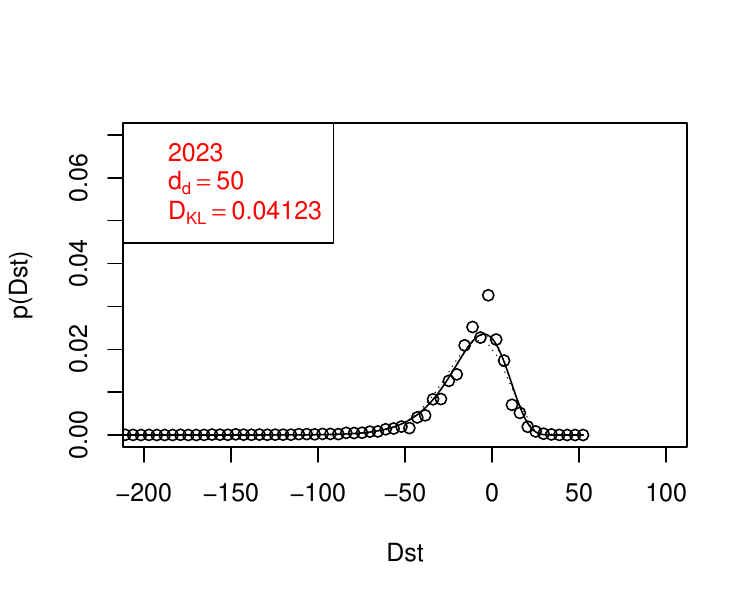}
\includegraphics[width=0.30\columnwidth]{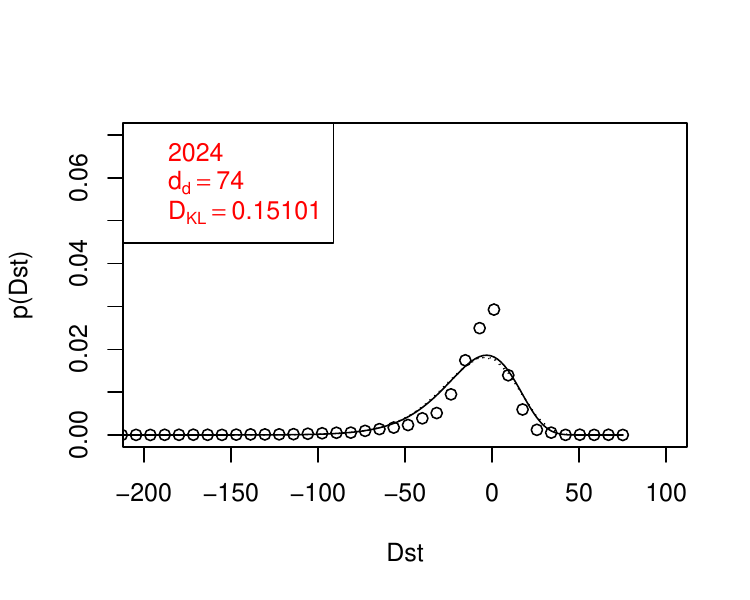}
\includegraphics[width=0.30\columnwidth]{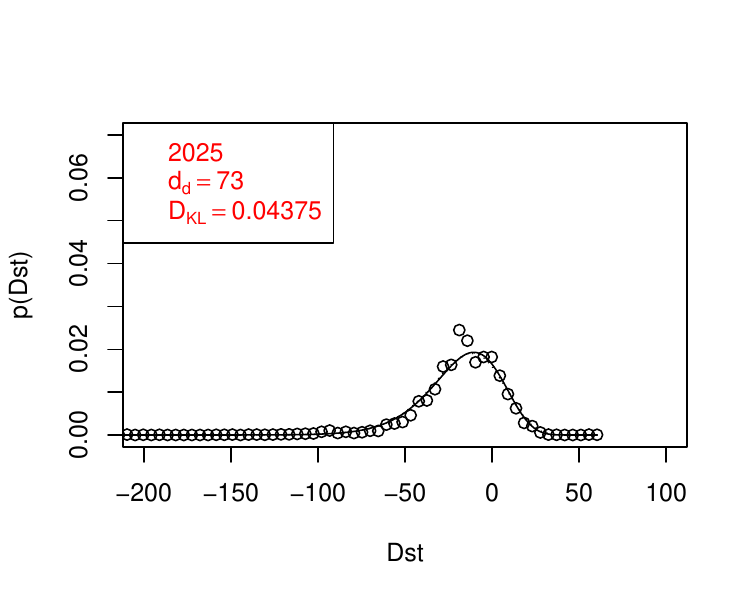}
\caption{Solar cycle 25, i.e. 2019 onwards.}
\label{fig-scycle25}
\end{figure}

These figures \ref{fig-scycle19} to \ref{fig-scycle25}
 show the fits for each year of ${\DS}$ data, 
 grouped into solar cycles.
The fitted distributions are typically a reasonable visual match
 to the data, 
 although there can 
 (a) be a slight staircasing of the ${\DS}$ histograms which affects the visual fit;
 or
 (b) a bump or distortion from the single-fit 
  hinting at a second contribution; 
 or 
 (c) a tendency for the data to be narrower and taller than the fit.
Legends in magenta and red indicate 
 larger KL divergences ${\DKL}$ than would be expected simply
 by sampling an ideally-matched distribution.

\clearpage
\section{Fit parameters}\label{S-fitparams}

Tables \ref{table-divisionsA} and \ref{table-divisionsB}
 show the full list of fit parameters
 and the KL divergence between that fit and the data.
Note that 
 the fitting parameters for the scaled ${\DS}'$ data, 
 are commonly denoted $\mu$ and $\sigma$ in the LN context.

\begin{table}
\begin{center}
\begin{tabular}{ || l | c | c | c | c || }
\hline
\hline
Year & ${\doff}$ & $\mu$ & $\sigma$ &  ${\DKL}$ \\
\hline
\hline
1957 & 57.2 & 4.284821 & 0.3915872 & 0.05305584\\
1958 & 58.6 & 4.28455 & 0.3748761 & 0.03837514\\
1959 & 83.4 & 4.652908 & 0.2728456 & 0.0357505\\
1960 & 53.7 & 4.363222 & 0.377088 & 0.0271085\\
1961 & 102.5 & 4.748058 & 0.1806916 & 0.09520211\\
1962 & 139.5 & 4.979537 & 0.1068383 & 0.02956233\\
1963 & 72.4 & 4.345889 & 0.2239046 & 0.06500677\\
1964 & 89.6 & 4.520687 & 0.1381726 & 0.04224836\\
1965 & 55 & 3.945637 & 0.2283002 & 0.06850486\\
1966 & 51.2 & 4.004731 & 0.2839695 & 0.05007002\\
1967 & 57.5 & 4.245373 & 0.2783986 & 0.08270811\\
1968 & 66.4 & 4.385524 & 0.2256496 & 0.06741474\\
1969 & 51.2 & 4.054301 & 0.3105854 & 0.04903814\\
1970 & 50.1 & 4.148936 & 0.2933913 & 0.04369205\\
1971 & 55.2 & 4.228535 & 0.2344861 & 0.03046975\\
1972 & 50.1 & 4.185309 & 0.2785964 & 0.04374615\\
1973 & 56.3 & 4.209086 & 0.2823537 & 0.03942486\\
1974 & 50.1 & 4.12161 & 0.2716726 & 0.04081375\\
1975 & 62.7 & 4.265365 & 0.1983713 & 0.03310324\\
1976 & 50.1 & 4.118932 & 0.2515551 & 0.06593655\\
1977 & 57.2 & 4.279945 & 0.2272421 & 0.04714848\\
1978 & 61.7 & 4.387262 & 0.2841941 & 0.0306295\\
1979 & 77.8 & 4.516026 & 0.2367179 & 0.03615956\\
1980 & 50.1 & 4.068423 & 0.3204775 & 0.01819519\\
1981 & 54.3 & 4.306256 & 0.3373557 & 0.03477094\\
1982 & 79.1 & 4.587446 & 0.289533 & 0.04235128\\
1983 & 59.4 & 4.295096 & 0.2868496 & 0.0406877\\
1984 & 80.2 & 4.560489 & 0.1987215 & 0.04020474\\
1985 & 50.1 & 4.156515 & 0.2469989 & 0.03290068\\
1986 & 50.1 & 4.141474 & 0.2842532 & 0.0442928\\
1987 & 63.5 & 4.299085 & 0.208592 & 0.03629948\\
1988 & 50.1 & 4.210383 & 0.2720476 & 0.04353944\\
1989 & 85.8 & 4.710357 & 0.2705119 & 0.04990936\\
1990 & 71.3 & 4.481166 & 0.2903335 & 0.0245874\\
1991 & 67 & 4.526908 & 0.321289 & 0.03992439\\
1992 & 59.6 & 4.326229 & 0.3232691 & 0.04347648\\
1993 & 69.1 & 4.436454 & 0.2296417 & 0.03422216\\
1994 & 51.5 & 4.248778 & 0.2762902 & 0.02071227\\
1995 & 50.1 & 4.167074 & 0.2665779 & 0.02772428\\
\hline
\hline
\end{tabular}
\end{center}
\caption{Data: 1957--1995, 
 solar cycles \#19 to \#22.}
\label{table-divisionsA}
\end{table}

\begin{table}
\begin{center}
\begin{tabular}{ || l | c |   c| c | c || }
\hline
\hline
Year & ${\doff}$ & $\mu$ & $\sigma$ &  ${\DKL}$ \\
\hline
\hline
1996 & 66.2 & 4.33092 & 0.1669386 & 0.02645945\\
1997 & 56.2 & 4.233275 & 0.2163797 & 0.04759897\\
1998 & 50.1 & 4.160909 & 0.2929323 & 0.05951495\\
1999 & 76.5 & 4.467646 & 0.2318523 & 0.02506829\\
2000 & 56.1 & 4.257113 & 0.3197647 & 0.05921238\\
2001 & 57.3 & 4.253191 & 0.3435982 & 0.09363659\\
2002 & 74.6 & 4.528896 & 0.2462567 & 0.05056975\\
2003 & 50.1 & 4.232002 & 0.2934625 & 0.06599569\\
2004 & 70.9 & 4.385431 & 0.2526929 & 0.1000152\\
2005 & 56.7 & 4.254593 & 0.2653719 & 0.04277711\\
2006 & 50.1 & 4.093935 & 0.2355232 & 0.02640821\\
2007 & 133 & 4.947335 & 0.07815317 & 0.04307636\\
2008 & 81.2 & 4.481638 & 0.1218884 & 0.04582673\\
2009 & 72 & 4.30986 & 0.1142158 & 0.05579793\\
2010 & 50.8 & 4.080181 & 0.1899734 & 0.07511392\\
2011 & 59.2 & 4.232382 & 0.2024039 & 0.07668242\\
2012 & 85.8 & 4.561836 & 0.1938002 & 0.08335846\\
2013 & 50.1 & 4.062829 & 0.2580175 & 0.03235881\\
2014 & 60.1 & 4.239995 & 0.2154245 & 0.03068618\\
2015 & 52.7 & 4.257098 & 0.2738056 & 0.05815063\\
2016 & 64.3 & 4.318567 & 0.2016269 & 0.04751752\\
2017 & 69.9 & 4.395265 & 0.1889373 & 0.05011629\\
2018 & 50.1 & 4.041052 & 0.2076956 & 0.03628705\\
2019 & 57.1 & 4.16479 & 0.1567391 & 0.05299092\\
2020 & 50.1 & 4.043392 & 0.1709047 & 0.03207812\\
2021 & 65 & 4.24079 & 0.1723404 & 0.04853644\\
2022 & 64.2 & 4.274026 & 0.2056214 & 0.03438393\\
2023 & 50.1 & 4.090136 & 0.2963234 & 0.04123291\\
2024 & 74.7 & 4.425386 & 0.2656423 & 0.1510054\\
2025 & 73.7 & 4.48998 & 0.2386214 & 0.04375393\\
\hline
\hline
\end{tabular}
\end{center}
\caption{Data: 1996--2025, 
 solar cycles \#23 to \#25}
\label{table-divisionsB}
\end{table}

\end{document}